\documentclass[10pt,a4paper]{article}
\usepackage[latin1]{inputenc}
\usepackage{latexsym}
\usepackage{amsbsy}
\usepackage{mathrsfs}
\usepackage{color}
\usepackage{psfrag}
\usepackage{enumerate}
\usepackage{amsmath,amssymb,calc,amsfonts}
\usepackage{latexsym}
\usepackage[colorlinks=true,linkcolor=blue,urlcolor=blue,citecolor=red]{hyperref}

\usepackage{amsfonts}
\usepackage{amssymb}

\usepackage{latexsym}
\usepackage{amsfonts}
\usepackage{t1enc}
\usepackage{times}
\usepackage{xmpmulti}
\usepackage{color}
\usepackage{bm}

\usepackage{graphicx}
\usepackage{subfigure} 
\usepackage{ifoddpage}
\usepackage{epstopdf} 
\usepackage{float}

\usepackage{setspace} 
\usepackage{courier} 
\usepackage{array,longtable} 
\usepackage{captdef}
\usepackage{slashbox}
\usepackage{caption}
\usepackage{comment}

\font\tenscr=rsfs10 scaled1100
\font\sevenscr=rsfs7 
\font\fivescr=rsfs5 
\skewchar\tenscr='177
\skewchar\sevenscr='177
\skewchar\fivescr='177
\newfam\scrfam
\textfont\scrfam=\tenscr
\scriptfont\scrfam=\sevenscr
\scriptscriptfont\scrfam=\fivescr

\begin{document}

\title{MALBEC: a new CUDA-C ray-tracer in General Relativity}
\date{\today}

\author{\small G. D. Quiroga$^\dagger$\\
\em \normalsize \em \small ${}^\dagger$ GIRG, Escuela de F\'isica, Universidad Industrial de Santander \\
\normalsize \em \small A. A. 678, Bucaramanga, Colombia\\
\normalsize \em \small gonzalo.quiroga1@correo.uis.edu.co
}
\maketitle

\begin{abstract}
A new CUDA-C code for tracing orbits around non-charged black holes is presented. This code, named MALBEC, take advantage of the graphic processing units
and the CUDA platform for tracking null and timelike test particles in Schwarzschild and Kerr. Also, a new general set of equations that describe the closed circular orbits of any timelike test particle in the equatorial plane is derived. These equations are extremely important in order to compare the analytical behavior of the orbits with the numerical results and verify the correct implementation of the Runge-Kutta algorithm in MALBEC. Finally, other numerical tests are performed, demonstrating that MALBEC is able to reproduce some well-known results in these metrics in a faster and more efficient way than a conventional CPU implementation.\\

\\

\end{abstract}

\maketitle

\section{Introduction}
The study and characterization of the geodesic motion around compact objects is extremely important in astrophysics. The orbits followed by test particles in the presence of a massive object can give information about the nature of the astrophysical source, also the dynamics around the compact sources like Neutron Stars (NS) or Black Holes (BHs) may be helpful to the gravitational wave observatories such LIGO \cite{aasi2015advanced}. For instance, during a binary black hole coalescence, in which one member of the binary is much more massive than the other. In this kind of system, it is possible to assume that the small body moves on a geodesic of the background Kerr spacetime \cite{sundararajan2010binary}, at least during the early-stages. This is, in the test-mass limit, the full nonlinear relativistic force law corresponds to a geodesic motion in a Kerr spacetime. Thus, one way to investigate the properties of such astrophysical sources is through the study of the geodesic motion in these backgrounds.

In the particular case of BHs, the geodesic structure around these compact sources was studied during decades. A detailed study of the geodesic motion around Kerr and
the Schwarzschild metric is summarized in the Chandrasekhar book \cite{chandrasekhar}, where timelike and null particles are analyzed, the stability of the trajectories is studied, and also the innermost stable circular orbits (ISCOs) are computed. In a recent work, the motion of classical spinning test particles in Schwarzschild and Kerr metrics was considered, and the ISCO was investigated \cite{jefremov2015innermost}. In the cited work, small-spin corrections for the ISCO parameters were found analytically. On the other hand, the circular motion of charged test particles in the gravitational field of a charged BH described by the Reissner-Nordstr\"om spacetime was studied. This makes it possible to develop techniques that allow to distinguish between black holes and naked singularities \cite{pugliese2011motion}.

Over the years, several analytical treatments have been carried out to track orbits around compact objects. For example, in the following reference \cite{levin2008periodic}, a system based on the Poincare approach is presented to classify orbits on the equatorial plane of rotating BH. In this mentioned article, the authors define a complete taxonomy of orbits that allows to establish a correspondence between periodic orbits and a set of rational numbers. On the other hand, the analytical solutions of bound timelike geodesic in Kerr were introduced in \cite{fujita2009analytical}. These solutions are given in terms of elliptic integrals using the Mino's method \cite{mino2003perturbative}, which consist in a perturbative approach to track an orbital evolution around a super massive black hole. Bound geodesic orbits around a Kerr black hole can be parametrized by three constants of motion which can be associated with three frequencies related to the radial, longitudinal and azimuthal motion. However, it was shown recently that these two ways to characterize bound geodesics are not in a one-to-one correspondence in the strong field regime \cite{warburton2013isofrequency}. That is, there is a region of the parameter space in which pairs of physically distinct orbits can have the same three frequencies.

The geodesic motion of test particles is not restricted only to theoretical analysis, the nonlinearity of the system of equations requires in many cases the use of numerical technics. Several computational codes can simulate and track the trajectories followed by test particles around BHs. For instance, the open source ray-tracer GYOTO developed by Vincent et al \cite{vincent2011gyoto}, or the GPU ray-tracer recently published by Daniel Kuchelmeister \cite{kuchelmeister2012gpu}. Both have a great handling of images and can integrate trajectories of test particles in vacuum solutions like Schwarzschild and Kerr, and in the case of GYOTO in numerical metrics as well. However, since our point of view, there several limitations in both implementations. For example, GYOTO has been developed using the standard CPU programming instead using the high performance of the actual graphics processing units (GPUs), which take advantage of the benefits offered by the CUDA platform developed by NVidia to accelerate the calculation of the orbits \cite{Nvidea}. On the other hand, the GPU ray-tracer, only can launch null geodesics in the Kerr and/or Schwarzschild metric, and it does not use the 3+1 formulation, which hinders the implementation of numerical metrics.

In this context arrived MALBEC, MALBEC is a new GPU code developed in CUDA-C for tracing orbits around compact objects in General Relativity (GR) that seeks to overcome the limitations of the previously mentioned implementations. Thus, in this work, we will introduce the first version of MALBEC, and we perform several validation tests. For this, we will analyze the motion of timelike particles around Schwarzschild and Kerr black holes. Now, from the mathematical aspect, MALBEC has been written using the 3+1 formulation of the GR and is able to handle timelike and null geodesic in Schwarzschild and Kerr spacetime. The use of the 3+1 formulation is important in order to bring future support to numerical metrics. Furthermore, in a future update, we will include extra analytical metrics as well as numerically computed, so the results presented in this article are not a limitation in our GPU implementation. From the technical issue, MALBEC implements the  Runge-Kutta method, labeled as RK4, to find the solution of the system of ordinary differential equations (ODEs) that describe the geodesic motion around the compact source. MALBEC has been written in the programming language C and uses the CUDA platform \cite{Nvidea} with a double precision arithmetic. Additionally, MALBEC  was tested on Linux and Windows, and can  work with a NVidia CUDA enabled GPU of computing capability 1.3 or higher, making use of the official NVidia NVCC compiler.

Since there are many technical details involved, it is better to outline the main ideas followed to perform the test of our code. In this way, the reader can have a broad picture without the technical complications. In order to perform the first validation of MALBEC, we will use the following procedure: first, we will derive the equations of motion for timelike and null particles by using the 3+1 formulation. These equations are written as a system of eight coupled ordinary differential equations. So, we obtain a general set of equations that allow to calculate the initial conditions of any circular orbits in the equatorial plane of Kerr and Schwarzschild at any position outside of the event horizon. From this approach, a simple set of equations for the orbital energy and angular momentum is obtained, which are used to construct the set of initial conditions that are evolved by MALBEC, thus, the closed orbits and the stability regions can be verified numerically showing the correct implementation of the code. In other words, we will verify that the numerical solution reflects the expected theoretical behavior. Then, we focus on the convergence and speed tests, where we show that the numerical method implemented converges to the expected numerical order, and also prove that our code is capable of running a bigger number of geodesics in a shorter time than an analogous CPU implementation.

This article is organized as follows. In the sec. \ref{Sec2}, some mathematical foundations are given, particularly the 3+1 decomposition of the GR is introduced, and the Kerr metric written in this formulation. In sec. \ref{Sec3}, the closed circular orbits in the equatorial plane of Kerr are derived, and the main equations of this work are presented. The special case of the circular orbits around a Schwarzschild black hole is treated in the sec. \ref{Sec4}. Then, the numerical simulations are performed in the sec. \ref{AN}. Also, in this section, we perform several tests in order to validate the efficient and correct execution of MALBEC. Finally, we closed the work given some final remarks and conclusions.

\section{Foundations}\label{Sec2}

In this section, we will introduce some mathematical foundations needed to develop our formalism. In particular, we will introduce the 3+1 decomposition of the GR which allows us to obtain the equations of motion followed by test particles around BHs, these equations of motion will be given as a system of ODEs.

In the 3+1 decomposition of the GR \cite{gourgoulhon}, the spacetime manifold $\mathcal{M}$ is assumed to be globally hyperbolic and to admit a foliation with spacelike hypersurfaces $\Sigma_{t}$, which are parametrized by the parameter $t \in \mathcal{R}: \mathcal{M} = \mathcal{R} \times \Sigma_{t}$ \cite{1979sgrr.work...83Y}. In this approach, the line element can be written as

\begin{equation}
ds^2 = -\alpha^2 dt^2 + \gamma_{ij}(dx^i + \beta^idt)(dx^j + \beta^jdt), \label{eq:ADM}
\end{equation}
where $\alpha$ defines the lapse of proper time $d\tau$, measured for those observers moving along the normal direction to the hypersurfaces $\mathbf{n} = -\alpha \mathbf{\nabla} t$, where $\mathbf{n}$ is a timelike future pointing 4-vector, $\mathbf{\beta}$ is a spatial vector called the {\it shift} vector and describes how the spatial coordinates shift when moving from the slice $\Sigma_{t}$ to another one $\Sigma_{t + \delta t}$, and $\gamma_{ij}$ are the components of the induced metric over each hypersurface, such that $\mathbf{\gamma}$ is a projector orthogonal to the timelike vector $\mathbf{n}$, {\it i.e.}, $\mathbf{\gamma} \cdot \mathbf{n} = 0 $.

Now, the equations of motion for a free particle in the generalized spacetime (\ref{eq:ADM}), can be obtained from the following Lagrangian \cite{gourgoulhon},
\begin{align}
{\cal L}&=-\frac{1}{2}\left( \alpha ^{2}-\sum_{i=1}^{3}\sum_{j=1}^{3}\gamma
_{ij}\beta ^{i}\beta ^{j}\right) \dot{t}^{2} \nonumber\\
&+\left(\sum_{i=1}^{3}\sum_{j=1}^{3}\gamma _{ij}\beta ^{j}\dot{x}^{i}\right)\dot{t} +\frac{1}{2}\sum_{i=1}^{3}\sum_{j=1}^{3}\gamma _{ij}\dot{x}^{i}\dot{x}^{j},
\end{align}
here the dot denotes differentiation with respect to an affine parameter $\tau$. Thus, following the canonical formulation for the geodesics \cite{1997ForPh..45..439V}, the equations of motion can be obtained by using the following expressions
\begin{equation}
\dot{x}^{\mu}  = \frac{\partial {\cal H}}{\partial p_{\mu}}, \qquad \dot{p}_{\mu} = -\frac{\partial {\cal H}}{\partial x^{\mu}}, \label{Ham}
\end{equation}
where
\begin{equation}
{\cal H} = \frac{1}{2}g^{\mu \nu}p_{\mu}p_{\nu}, \qquad p_{\mu} = \frac{\partial {\cal L}}{\partial \dot{x}^{\mu}},
\end{equation}
are the Hamiltonian and the four momentum, respectively. It is worth mentioning that the energy is only derived from the kinetic energy in such way that ${\cal H}={\cal L}$. In the case of charged particles this relation is not satisfied, that is  ${\cal H}\neq {\cal L}$. Furthermore, the geodesic equation for a timelike particle, wrote in the 3+1 formulation, obey the following constraint,
\begin{equation}\label{geo}
2{\cal L}=-1
\end{equation}

Throughout this paper, we assume geometrized units, that is, $G=c=1$. Also, we introduce the Kerr line element in Boyer-Lindquist (BL) coordinates $x^{\alpha} = (t,r,\theta,\phi)$ as \cite{chandrasekhar},
\begin{align}
\nonumber ds^2 &= -\left(\frac{\Delta - a^2 \sin^2\theta}{\rho^2}\right)dt^2 - 4\frac{aMr\sin^2\theta}{\rho^2}dtd\phi \\
&+ \frac{\rho^2}{\Delta}dr^2 + \rho^2 d\theta^2 + \frac{\Sigma}{\rho^2} \sin^2\theta d\phi^2, \label{eq:kerr}
\end{align}
with the usual definitions
\begin{align*}
\rho^2 &= r^2 + a^2\cos^2\theta, \\
\Delta &= r^2 - 2Mr + a^2, \\
\Sigma &= \left( r^2 + a^2\right)^2 - a^2\Delta \sin^2 \theta.
\end{align*}
Here $M$ is the mass of the BH, and $a$ the black hole angular momentum per unit mass $ (J/M) $, note that in the case of the Schwarzschild metric the parameter $a=0$. Additionally, we will investigate only the case $a\leq M$, since for $a>M$ the Kerr metric describes a naked singularity. Furthermore, the coordinate transformation between BL  coordinates $(r , \theta, \phi)$ and the Cartesian $(x, y, z)$ is given by
\begin{align} \label{coord}
x&=\sqrt{r^2+a^2}\sin \theta \cos \phi , \nonumber\\
y&=\sqrt{r^2+a^2}\sin \theta \sin \phi , \\
z&=r \cos \theta. \nonumber
\end{align}
Now, by comparing equations (\ref{eq:ADM}) and (\ref{eq:kerr}), we get
\begin{equation}
\alpha =  \sqrt{\frac{\rho^2 \Delta}{\Sigma} }, \quad
\beta^{i} = \left( 0,0, -\frac{2 M a r}{\Sigma}\right), \quad
\gamma_{ij}  = \text{diag}\left[ \frac{\rho^2}{\Delta}, \rho^2, \frac{\Sigma}{\rho^2}\sin^{2}\theta \right],
\end{equation}
in this way the Kerr metric is expressed in terms of the lapse function $\alpha$, the shift vector $\beta^{i}$, and the components of the induced metric $\gamma_{ij}$.

\section{Closed circular orbits it the equatorial plane} \label{Sec3}

In this section, we will obtain the equations of the closed circular orbits needed to test MALBEC. These orbits correspond to timelike test particles moving in closed circular trajectories in the equatorial plane of a non-charged black hole. The equatorial plane is the plane defined by the condition $z=0$, or in BL coordinates $\theta=\pi/2$. Now, the restriction $\theta=\pi/2=const.$ imposes the following condition $\dot \theta=0$, thus, the ODEs system coming from the eqs. (\ref{coord}) are reduced to the following,
\begin{align}
\dot{t} &=\frac{-(2Ma^{2}+a^{2}r+r^{3})p_{t}+2aMp_{\phi }}{%
r(2Mr-a^{2}-r^{2})}, \\
\dot{r} &=-\frac{(2Mr-a^{2}-r^{2})}{r^{2}}p_{r}, \label{dotr}\\
\dot{\phi} &=\frac{-2Map_{t}+(2M-r)p_{\phi }}{r(2Mr-a^{2}-r^{2})},\label{dotphi}\\
\dot{p}_{r} &=-\frac{(Mr-a^{2})}{r^{3}}p_{r}^{2}+\frac{M(4Ma^{2}r-a^{4}-2a^{2}r^{2}-r^{4})}{r^{2}(2Mr-a^{2}-r^{2})^{2}}p_{t}^{2} \label{dotpr} \\
&-\frac{2Ma(4Mr-a^{2}-3r^{2})}{r^{2}(2Mr-a^{2}-r^{2})^{2}}p_{t}p_{\phi }+\frac{(4M^2r-Ma^{2}-4Mr^{2}+r^{3})}{r^{2}(2Mr-a^{2}-r^{2})^{2}}p_{\phi }^{2}, \nonumber \\
\dot{\theta} &= \dot{p}_{t} =\dot{p}_{\theta } = \dot{p}_{\phi } =0,
\end{align}
In order to get these circular closed trajectories around Kerr, it is necessary to keep the radial coordinate constant, and also the radial canonical momentum, i. e.,
\begin{equation}
\dot r=0, \qquad p_r=0.
\end{equation}
Note that these two conditions are compatible with eq. (\ref{dotr}), also the last condition implies that $\dot{p}_r=0$. Now, to guarantee that $p_r = 0$ in eq. (\ref{dotpr}), we will set to zero the l.h.s of (\ref{dotpr}), and we will solve for $p_r^2$, from where we can write,
\begin{align}
p_r^2&=\frac{r(4M^{2}r-Ma^{2}-4Mr^{2}+r^{3})}{%
(Mr-a^{2})(2Mr-a^{2}-r^{2})^{2}}p_{\phi }^{2} \label{eq1}\\
&-\frac{2raM(4Mr-a^{2}-3r^{2})}{(Mr-a^{2})(2Mr-a^{2}-r^{2})^{2}}%
p_{t}p_{\phi } \nonumber \\
&+\frac{rM(4Ma^{2}r-a^{4}-2a^{2}r^{2}-r^{4})}{%
(Mr-a^{2})(2Mr-a^{2}-r^{2})^{2}}p_{t}^{2} \nonumber
\end{align}
Then, we introduce the last equation into the geodesic equation coming from (\ref{geo}) to get the following,
\begin{align}
-1 &=-\frac{(M-r)(2Mr+a^{2}-r^{2})}{r(Mr-a^{2})(2Mr-a^{2}-r^{2})}p_{\phi
}^{2} \label{eq2}\\
&+\frac{2Ma(2Mr+a^{2}-3r^{2})}{r(Mr-a^{2})(2Mr-a^{2}-r^{2})}p_{t}p_{\phi }
\nonumber \\
&-\frac{(2M^{2}a^{2}r+Ma^{4}-3Ma^{2}r^{2}-2Mr^{4}+a^{4}r+a^{2}r^{3})}{%
r(Mr-a^{2})(2Mr-a^{2}-r^{2})}p_{t}^{2}\nonumber
\end{align}
Finally, we have obtained a system of two algebraic equations given by (\ref{eq2}), and by the r.h.s of (\ref{eq1}) equal to zero, i.e. $p_r^2=0$. On the other hand, it is very usual to associate the orbital energy $E$ and the angular momentum about the normal axis to the equatorial plane $L$, to the canonical momenta $p_t$ and $p_\phi$ respectively. Now, solving the system given by eqs. (\ref{eq1}) and (\ref{eq2}) for positive energies $E>0$, and writing the equations in terms of $E$ and $L$, we find the following,
\begin{align}
E^{\pm } &=\left\vert \frac{4M^{2}r-Ma^{2}-4Mr^{2}+r^{3}}{\sqrt{R\pm
2M^{3/2}r^{3/2}(2Mr-a^{2}-r^{2})a}}\right\vert \label{energia} \\
L^{\pm } &=\pm \left\vert \frac{%
Ma(4Mr-a^{2}-3r^{2})+r^{3/2}(2Mr-a^{2}-r^{2})M^{1/2}}{\sqrt{R\pm
2M^{3/2}r^{3/2}(2Mr-a^{2}-r^{2})a}}\right\vert \label{momento}
\end{align}
Here the superscript "$^+$" represent the co-rotating orbit, while "$^-$" the counter-rotating, also the symbol $R$ is introduced for shorts, and it is given by,
\begin{equation}
R=r^{2}(-12M^{3}r+5M^{2}a^{2}+16M^{2}r^{2}-3Ma^{2}r-7Mr^{3}+r^{4}).
\end{equation}
Note the circular trajectories exist only if $r$ takes values such that the factor inside of the square root be positive. On the other hand, the minimal in the energy for the co-rotating, or the counter-rotating orbit, can be found from the first derivative of eq. (\ref{energia}), in the following way,
\begin{equation}\label{ISCO}
\frac{E^{\pm}}{dr} =0,
\end{equation}
then, we solve (\ref{ISCO}) for $r$, the roots of interest must meet simultaneously the following conditions: $r$ must be a real value, should be located outside of the event horizon $r_+$ of the BH, that is $r \geq r_{+}$ with $r_{+}=M+\sqrt{M^2-a^2}$ \cite{poisson2004relativist}. Also, the test particle for this radius should have a positive energy and real angular momentum. Thus, this root will correspond to the ``Innermost Stable Circular Orbits'' (ISCO), which is the smallest orbit where the test particle can stably stable circular orbits around a BH. So, in our approach, the ISCO is given by,
\begin{align}
r_{isco}^{\pm } &=3M+\frac{1}{2}\sqrt{A+B+C} \label{iscokerr} \\
&\mp \sqrt{-A-B+2C+\frac{224a^{2}M+1728M(a^{2}-6M^{2})}{4\sqrt{A+B+C}}}, \nonumber
\end{align}
where
\begin{align*}
A &=\frac{4(a^{4}-10a^{2}M^{2}+9M^{4})}{(a^{6}+17a^{4}M^{2}-45a^{2}M^{4}+27M^{6}+8\sqrt{a^{10}M^{2}-2a^{8}M^{4}+a^{6}M^{6}})^{1/3}}, \\
B &=4(a^{6}+17a^{4}M^{2}-45a^{2}M^{4}+27M^{6}+8\sqrt{a^{10}M^{2}-2a^{8}M^{4}+a^{6}M^{6}})^{1/3}, \\
C &=36M^{2}+4(a^{2}-6M^{2}),
\end{align*}
with $0\leq a < M$. Furthermore, the ISCO for the extreme Kerr can be found taking the limit $a \rightarrow M$ in eq. (\ref{iscokerr}), doing this we recover the well know results \cite{bardeen1972rotating},
\begin{align}
r_{isco}^{+} &=M, \quad L_{isco}^+ = \frac{2}{\sqrt{3}}M, \quad E_{isco}^+=\frac{1}{3}, \label{exKerr+} \\
r_{isco}^{-} &=9M, \quad L_{isco}^-=-\frac{22}{3\sqrt{3}}M, \quad E_{isco}^-=\frac{5}{3\sqrt{3}}, \label{exKerr-}
\end{align}
where $L^{\pm}$ and $E^{\pm}$ are the energy and the angular momentum associated to the ISCO radius, which is computed from the eqs. (\ref{energia}) and (\ref{momento}). In order to find the energy and the angular momentum associate with the co-rotating orbit in the extreme Kerr, $L_{isco}^+,E_{isco}^+$, must be taken the right-hand limit $r \rightarrow M^+$ since the radius of the orbit coincides with the event horizon. Finally, in the sec. \ref{AN}, we will use these equations to compute several ISCOs for Kerr.

On the other hand, in a closed circular orbit, the geodesics will describe a harmonic motion around the black hole. For those trajectories, the cartesian coordinates $(x,y)$ will obey,
\begin{align}\label{mas}
  x &= r \cos(\omega \tau +\alpha), \\
  y &= r \sin(\omega \tau +\alpha),
\end{align}
where $\tau$ is the proper time of the test particle, $\alpha$ is an initial phase, and $r$ is the constant radius of the circular orbit. Now, the angular frequency $\omega$ can be found comparing eqs. (\ref{mas}) with (\ref{coord}) assuming $\theta=\pi/2$, thus we can find the following equation,
\begin{equation}
\omega \tau + \alpha = \phi.
\end{equation}
Now, by taking one derivative and using (\ref{dotphi}) we get,
\begin{equation}\label{omega}
\omega=\dot{\phi}=\frac{-2Map_{t}+(2M-r)p_{\phi }}{r(2Mr-a^{2}-r^{2})},
\end{equation}
so, the period of the motion will be given by,
\begin{equation}
T=\frac{2\pi}{\omega}=\frac{2 \pi r(2Mr-a^{2}-r^{2})}{-2Map_{t}+(2M-r)p_{\phi }} \label{periodo}
\end{equation}
thus, the time taken by the particle to complete a lap around the Kerr BH depends on the radius, the energy, and the angular momentum.

\newpage

\section{Circular orbits around Schwarzschild black hole} \label{Sec4}
In this section, we will apply our previous results in a spherically symmetric static black hole, which is given by the Schwarzschild metric. The Schwarzschild metric is a vacuum solution to the Einstein field equations that describes the gravitational field outside of a non-rotating massive spherically-symmetric object. It is possible to reduce the Kerr solution to the Schwarzschild ones by setting the parameter $a=0$, thus the equations of motion for a timelike test particle are simpler than Kerr, and also can be obtained from (\ref{Ham}). These equations can be written as follows \cite{chandrasekhar},
\begin{align}
&\dot{t}= \frac{p_t}{1-\frac{2 M}{r}} \qquad \dot{r}= \left(1-\frac{2 M}{r}\right){p_r} \\
& \dot{\theta}= \frac{p_{\theta}}{r^2} \qquad \dot{\phi}=\frac{p_{\phi}}{r^2 \sin^2 \theta} \\
& \dot{p}_r=-\frac{M p_t^2}{(-r+2M)^2}-\frac{M p_r^2}{r^2}+\frac{p_{\theta}^2}{r^3}+\frac{p_\phi^2}{r^3\sin^2\theta} \\
& \dot{p}_t=0 \qquad \dot{p}_{\theta}=\frac{p_{\phi}^2 \cos \theta}{r^2 \sin^3 \theta} \qquad \dot{p}_\phi=0
\end{align}
In the equatorial plane, the circular orbits around the BH are determined by the conditions,
\begin{equation}\label{IC1}
\theta = \pi/2, \quad r=const.,
\end{equation}
which implies that $\dot \theta=0 \rightarrow p_{\theta}=0$, $\dot r=0 \rightarrow p_r=0$, and $\dot p_r=0$. Now, the equations for the energy and the angular momentum can be obtained from the eqs. (\ref{energia}) and (\ref{momento}) just imposing the condition $a=0$, from where we can write,
\begin{align}
E^2 &= \frac{4M^2-4Mr+r^2}{r(r-3M)}, \label{COrbit}  \\
L^2 &= \frac{Mr^2}{r-3M}. \label{COrbit2}
\end{align}
Note the energy must be a positive quantity, so one can observe that circular orbits for a timelike particle could not exist for $r\leq 3M$. Now, the minimal in the energy is found by taking a derivative to eq. (\ref{COrbit}) and solving for $r$ as follows,
\begin{equation}
\frac{d E}{dr}=\frac{M(6M-r)}{2r^{3/2}(r-3M)^{3/2}}=0.
\end{equation}
Thus, for $r=6M$ we get the critical point which corresponds to the Schwarzschild ISCO. Now, evaluating the equations (\ref{COrbit}) and (\ref{COrbit2}) in $r=6M$ we get
\begin{equation}\label{isco}
r=6M \qquad E=\sqrt{\frac{8}{9}} \qquad L=\sqrt{12}M.
\end{equation}
The values $r=3M$ and $r=6M$ defines two ranges for the radius where the orbits will have different behaviors and stabilities \cite{chandrasekhar},
\begin{equation}
6M<r \text{(stable)}<\infty, \qquad \text{and} \qquad 3M \leq r \text{(unstable)} \leq 6M.
\end{equation}
The circular orbits of the larger radius, where the minimum of the energy is located, will be stable in contrast to the circular orbit of the smallest radius where the energy is undefined \cite{chandrasekhar}.

Now, in several applications, it is quite convenient to assume the radius of the orbit as proportional to the mass, i.e. $r=k M$ with $k>0$. In general, if we write $r=k M$ for Schwarzschild the constant must satisfy $k>3$ in order to get closed orbits, thus we can write the following equations,
\begin{equation}\label{gralorbits}
r=k M \qquad p_t=\sqrt{\frac{k^2-4k+4}{k(k-3)}} \qquad p_\phi=\frac{k M}{\sqrt{k-3}}
\end{equation}
Finally, putting $a=0$ in eq. \ref{periodo} we get the period of the circular motion,
\begin{equation}
T=\frac{2\pi}{\omega}=\frac{2 \pi r^2}{p_\phi},
\end{equation}
thus, for a non-rotating BH, the time to complete a lap around the singularity do not depends on energy, just depends on the radius and the angular momentum.

\section{CUDA-C implementation for computing orbits in General Relativity}\label{sec:GPGPU}

In 2006, CUDA (Compute Unified Device Architecture) was introduced by NVidia, this new platform has opened the possibility of using NVidia GPUs to parallel computing. CUDA leverages the parallel compute engine in video cards to solve many complex computational problems in a more efficient way than on a CPU. Also, the Nvidia tool-kit comes with a software environment that allows to use C as a high-level programming language \cite{nvidia2011nvidia}.

The CUDA platform has many benefits for paralleling computing, benefits such as the existence of a shared memory (i.e. a memory area to be shared between threads), faster loads of data between GPU and CPU, full support for integer, bitwise operations and scattered reads. Despite CUDA has many advantages over other types of computing systems, there are some limitations that must be taken into account during its implementation. For example, in single precision NaNs are not supported, you can not use pointers to functions or functions with variable parameters. Also, for reasons of efficiency, the threads must be launched in groups of at least 32 with thousands of threads in total, but the biggest drawback, is a bottleneck between the CPU and the GPU by bandwidths and latencies buses. This bottleneck may affect the rate of data transfer CPU-GPU and thus the efficiency in the execution of the kernels.

Malbec, is a new CUDA-C code for tracing orbits around compact sources, this code takes advantage of the graphic processing units and the CUDA platform in order to track the geodesic motion of timelike and null test particles in curved spacetimes. The code is divided into several source files and headers located in different folders as one can see in the following diagram.
\begin{figure}[H]
\centering
\includegraphics[scale=0.4]{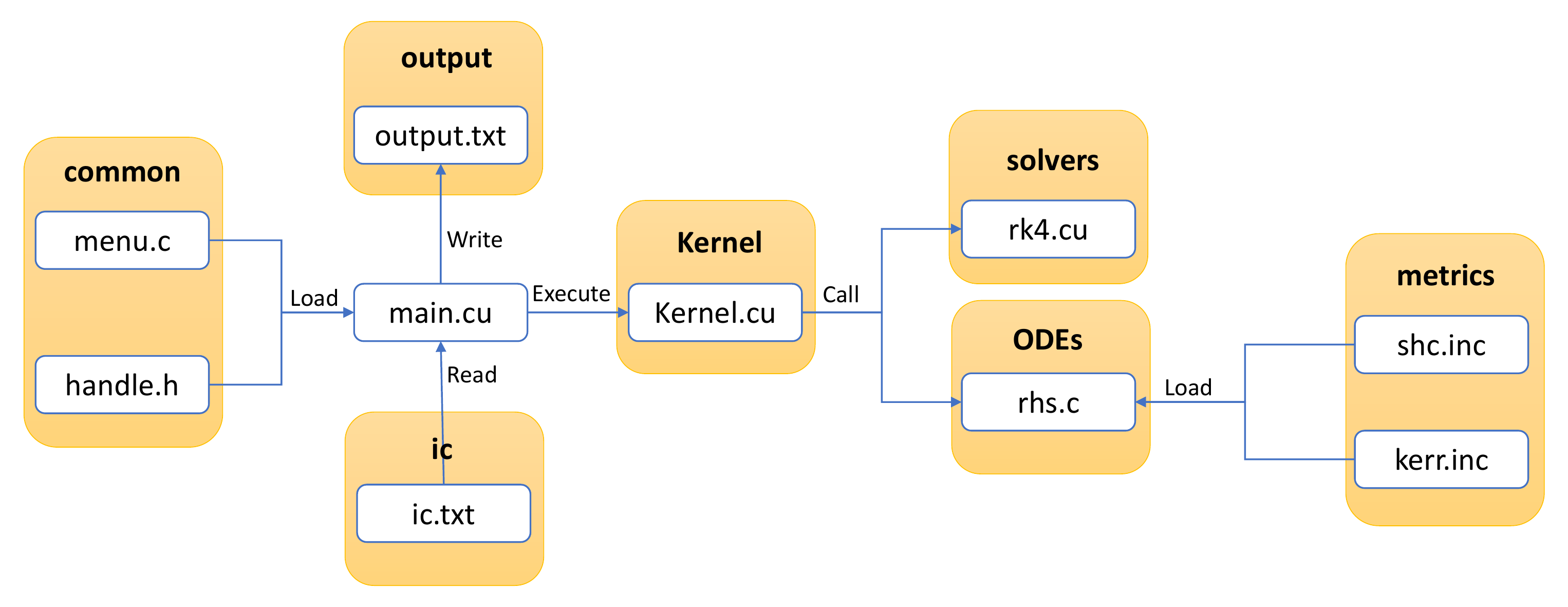}
\caption{\emph{MALBEC code structure. The orange blocks represent the folders where the files are located, for simplicity some headers are not included in the diagram.}}
\label{fig1}
\end{figure}
The main.cu is the main file of the code, this introduces some spacetime parameters such as the BH rotation, the number of initial conditions, and controls all the necessary processes during the execution. The code can be compiled with the standard nvcc, e.g. in a UNIX system like Ubuntu, the command line will be \texttt{nvcc main.cu -o malbec.out}. In the main file, the host and device input/output are defined, also, following the standard CUDA programming, the GPU is denoted as device and the CPU as host. The memory for each vector on the GPU and CPU is allocated, and the initial conditions are read. Moreover, the initial conditions are stored in a text file as an array of $8\times N$, where $N$ is the desired number of independent initial conditions or systems.
\begin{table}[H]
\centering
\begin{tabular}{c c c c c c c c }
 $t$ & $r$ & $\theta$ & $\phi$ & $p_t$ & $p_r$ & $p_{\theta}$ & $p_{\phi}$ \\
0.0000&	6.0000&	1.5708&	0.0000&	0.9428&	0.0000&	0.0000&	3.4641\\
0.0000&	3.5000&	1.5708&	0.0000&	1.1339&	0.0000&	0.0000&	4.9497\\
0.0000&	4.0000&	1.5708&	0.0000&	1.0000&	0.0000&	0.0000&	4.0000 \\
$\vdots$ & $\vdots$ &$\vdots$ &$\vdots$ & $\vdots$ & $\vdots$ &$\vdots$ &$\vdots$
\end{tabular}
\end{table}
Each row corresponds to the initial conditions of a test particle ordered as shown in the previous diagram. Note that each initial condition is a double precision number, however, for simplicity we show four digits.

Additionally, the ODEs system is composed by eight differential equations which are hosted in the ODEs folder, the rhs.c handles the sets of equations. In the current version, MALBEC includes the geodesic equations in Kerr and Schwarzschild spacetime. However, in a future update, other spacetimes will be included and also the code capabilities will be extended to handle numerical metrics. MALBEC use a GPU implementation of the traditional fourth order Runge-Kutta (RK4) method, others method of the RK family will be included soon. The use of the Runge-Kutta family is a well extended practice in order to solve a system of ODEs, a complete description of the implicit and explicit RK solvers can be found in ref. \cite{butcher2016numerical}. Additionally, the output file is an ASCII text, where all the generated data during the integration is written, this file will be stored in the output folder in the following format,

\begin{table}[H]
\centering
\begin{tabular}{c c c c c l}
$\tau$ & $t$ & $x$ & $y$ & $z$ \\
0.0000&	0.0000&	6.0000&	0.0000&	0.0000&$\rightarrow$ i.c. for the 1$^{st}$ test particle\\
0.0000&	0.0000&	3.5000&	0.0000&	0.0000&$\rightarrow$ i.c. for the 2$^{nd}$ test particle\\
0.1000&	0.1414&	5.9997&	0.0577&	0.0000&$\rightarrow$ 1$^{st}$ test particle in $\tau=h$\\
0.1000&	0.2646&	3.4971&	0.1414&	0.0000&$\rightarrow$ 2$^{nd}$ test particle in $\tau=h$\\
0.2000&	0.2828&	5.9989&	0.1155&	0.0000&$\rightarrow$ 1$^{st}$ test particle in $\tau=2h$\\
0.2000&	0.5292&	3.4886&	0.2825&	0.0000&$\rightarrow$ 2$^{nd}$ test particle in $\tau=2h$\\
$\vdots$ & $\vdots$ &$\vdots$ &$\vdots$ & $\vdots$ & $\vdots$
\end{tabular}
\end{table}
The previous example shows the output file for two geodesics in Schwarzschild. The output.txt file is printed following the same particle order in the ic.txt file, in addition, the numerical solutions are stored in Cartesian coordinates.

In CUDA, the threads are grouped into blocks, and the blocks are grouped into grids. All threads running in the same block share the fast shared memory, these threads can exchange  data using this shared memory. It is convenient to understand the restrictions on the kernel and the GPU in which the different threads are running, because the proper choice of the block size can affect the code performance. One of the keys to improve the performance is to keep the multiprocessors on the device as busy as possible. To implement this idea, the maximum occupancy criterion is used in MALBEC. The occupancy is the ratio of the number of active warps per multiprocessor to the maximum number of warps that can be active on the multiprocessor at once. Higher occupancy does not always translates into higher performance, there is a point above which additional occupancy does not improve performance. However, low occupancy always interferes with the ability to hide memory latency resulting in performance degradation, according to the CUDA-C best practices guide \cite{cudabest}.

The kernel.cu file is responsible for calling the integrator method hosted in the file rk4.cu and the rhs.c file, which loads the system of differential equations. Now, the integration goes on until one of the following stop conditions are fulfilled.
\begin{itemize}
\item The final evolution time is reached.
\item The particle approaches to the event horizon, i.e. the radial coordinate $r \leq 1+\sqrt{1-a^2}$.
\end{itemize}
The first stop condition controls the global iterations, that is, the main loop of all threads in the main file. This condition stops the evolution of all initial conditions when final time set by the user is reached. On the other hand, the second stop condition is checked individually for each thread. This verification is written inside of the rk4.cu file and stops the evolution of the particular thread when the test particle reach to the event horizon.

Finally, there are an auxiliary header and an extra file located in the common folder, these files contains some functions definition needed to check the right data transfer to the device, and also to display the MALBEC option menu. Finally, the numerical implementation made in MALBEC allows to solve a large set of system independently. Currently, MALBEC continues under development, however, this first release v0.5.0 is available and free to download from: \textcolor[rgb]{0.00,0.07,1.00}{https://github.com/GonzaQuiro/MALBEC}.

\section{Numerical analysis} \label{AN}
In this section, we will perform several simulations of timelike orbits, these trajectories are numerically evolved using MALBEC. As mentioned above, the goal in this work is to introduce MALBEC into the community and perform several tests to ensure proper execution of our GPU implementation, and also measures its performance. For this, we start by studying the circular motion of timelike test particles around Schwarzschild and Kerr BH using the standard RK4 method with a step-size $h=0.001$. The initial conditions are computed by eqs. (\ref{energia}) and (\ref{momento}) assuming a unit mass as is usual in the numerical practices. Then, at the end of this section, we focus on the calculation of the convergence rates, we also make a performance test comparing our GPU scheme against an analogous implementation in CPU.

\subsection{Schwarzschild black hole}
\begin{figure}[H]
  \checkoddpage
  \edef\side{\ifoddpage l\else r\fi}%
  \makebox[\textwidth][\side]{%
    \begin{minipage}[c]{0.5\textwidth}
      \centering
      \includegraphics[width=6.6cm]{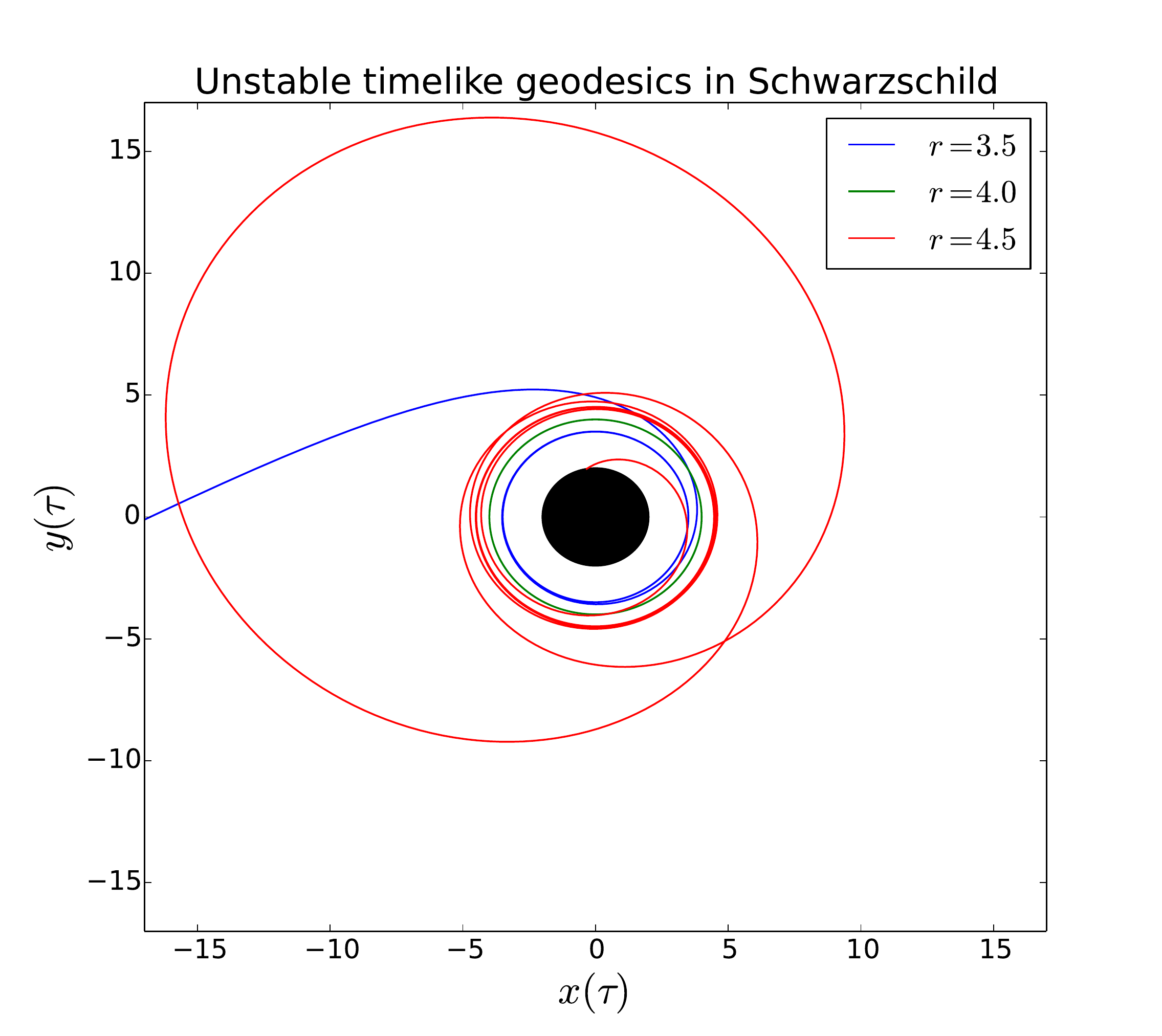}
    \end{minipage}%
    \hfill
    \begin{minipage}[c]{0.5\textwidth}
      \centering
      \includegraphics[width=6.6cm]{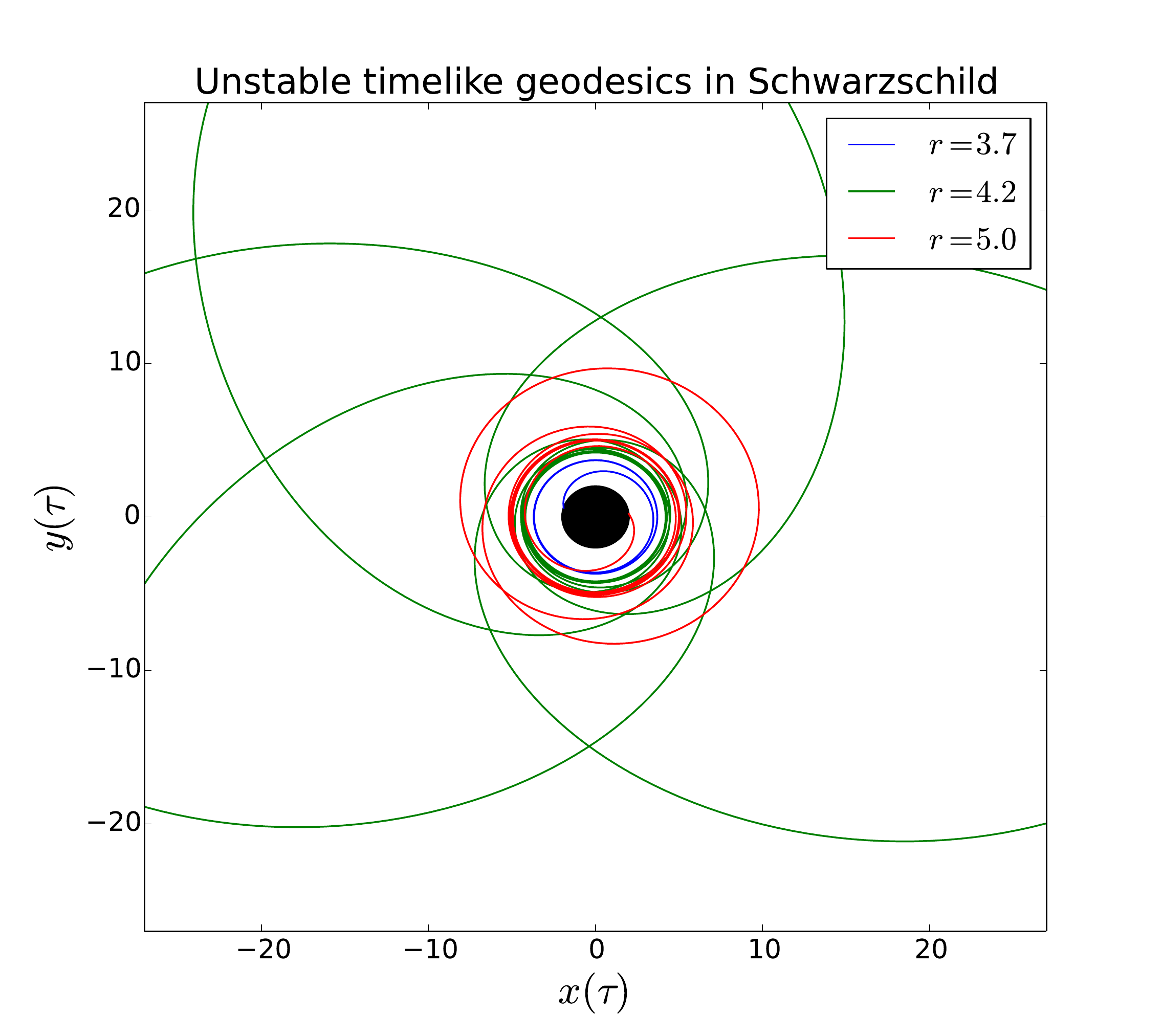}
    \end{minipage}%
  }
\caption{These figures show some unstable orbits around Schwarzschild. Since these orbits are unstable, will not keep in its circular closed trajectories. }
\label{Fig1}
\end{figure}

\begin{figure}[H]
  \checkoddpage
  \edef\side{\ifoddpage l\else r\fi}%
  \makebox[\textwidth][\side]{%
    \begin{minipage}[c]{0.5\textwidth}
      \centering
      \includegraphics[width=6.6cm]{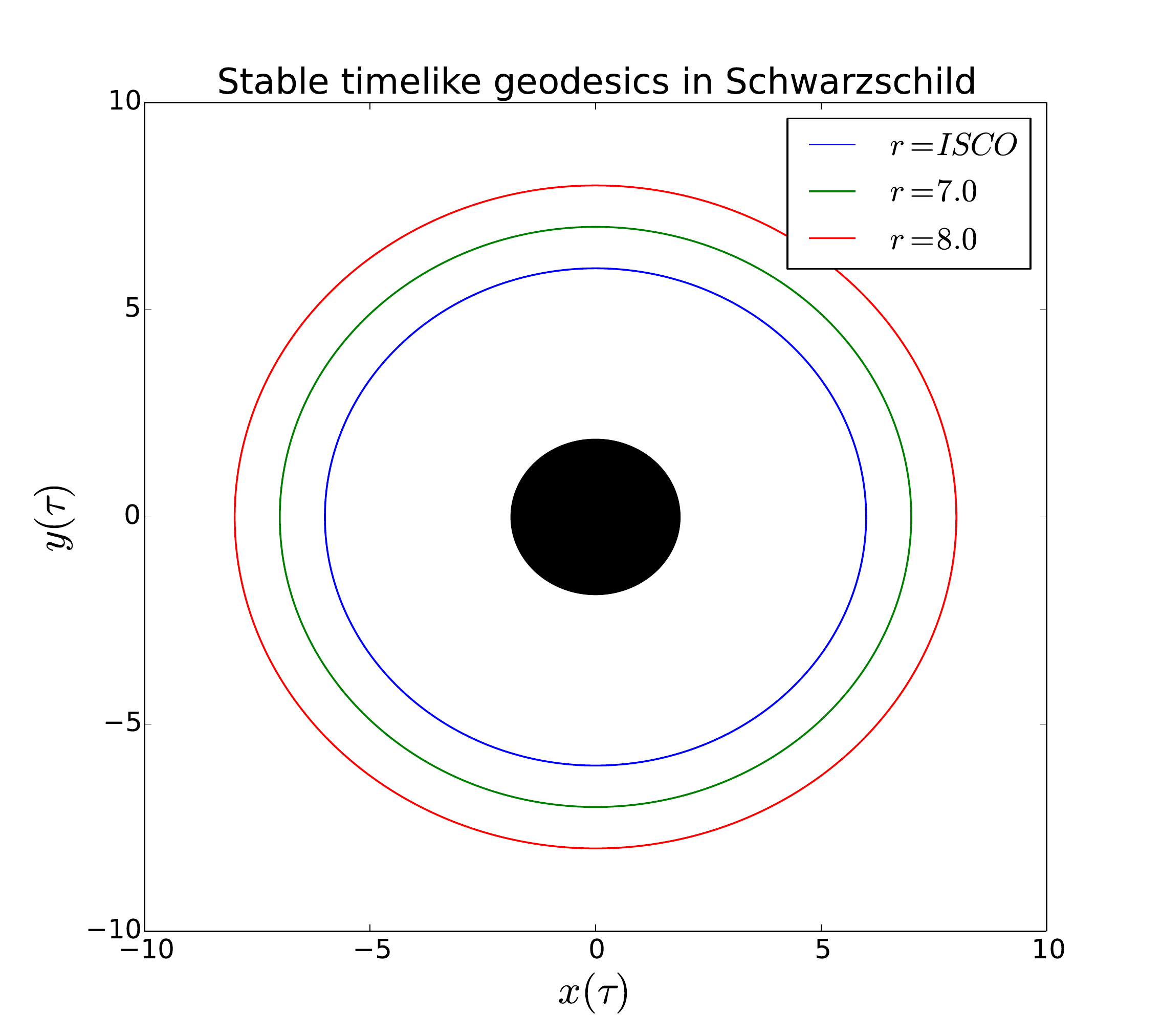}
    \end{minipage}%
    \hfill
    \begin{minipage}[c]{0.5\textwidth}
      \centering
      \includegraphics[width=6.6cm]{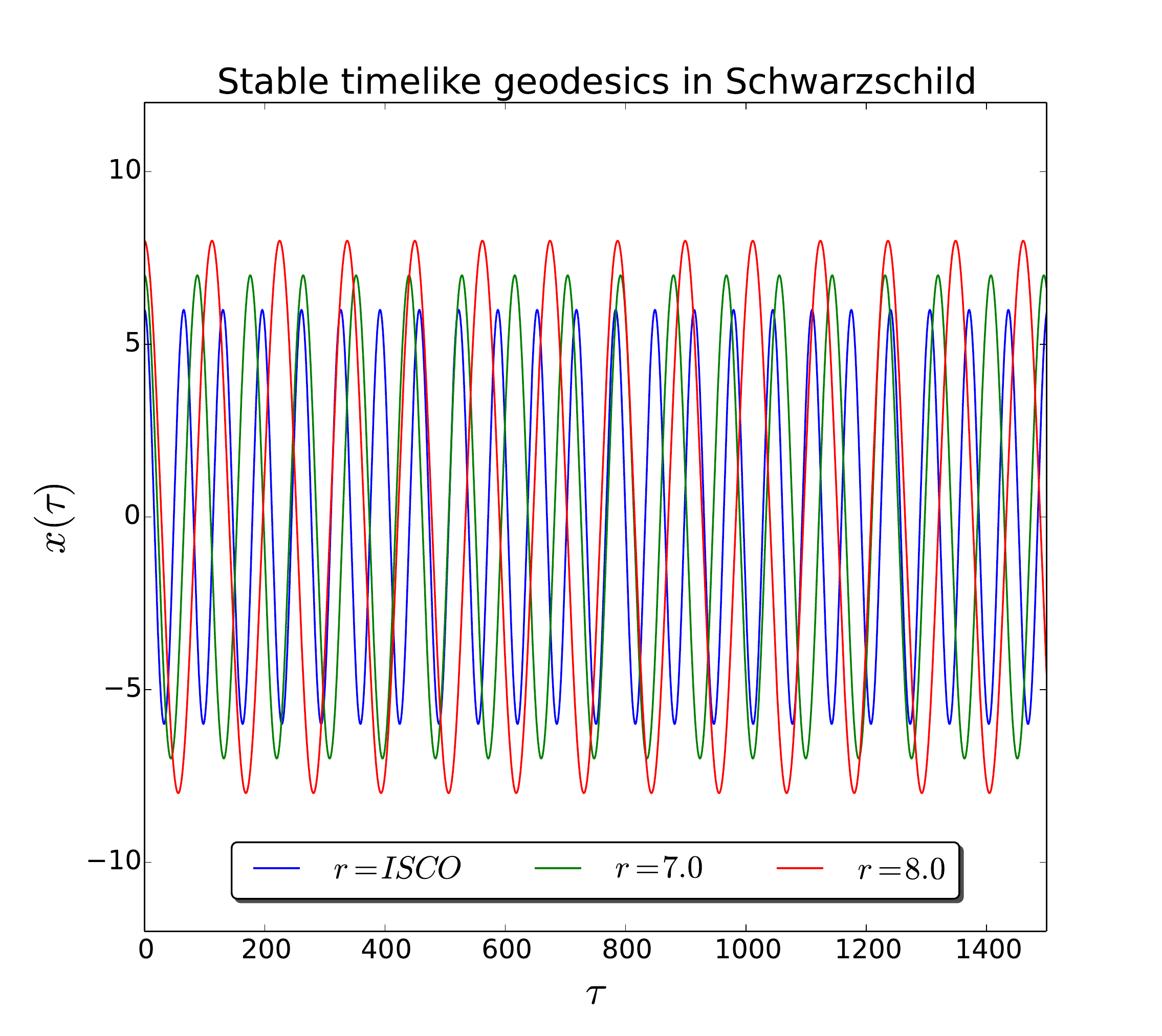}
    \end{minipage}%
  }
\caption{Some stable trajectories around the black hole, in the right figure each oscillation of the $x$ coordinate represent a lap around the compact source.}
\label{Fig2}
\end{figure}

\begin{figure}[H]
  \checkoddpage
  \edef\side{\ifoddpage l\else r\fi}%
  \makebox[\textwidth][\side]{%
    \begin{minipage}[c]{0.5\textwidth}
      \centering
      \includegraphics[width=6.6cm]{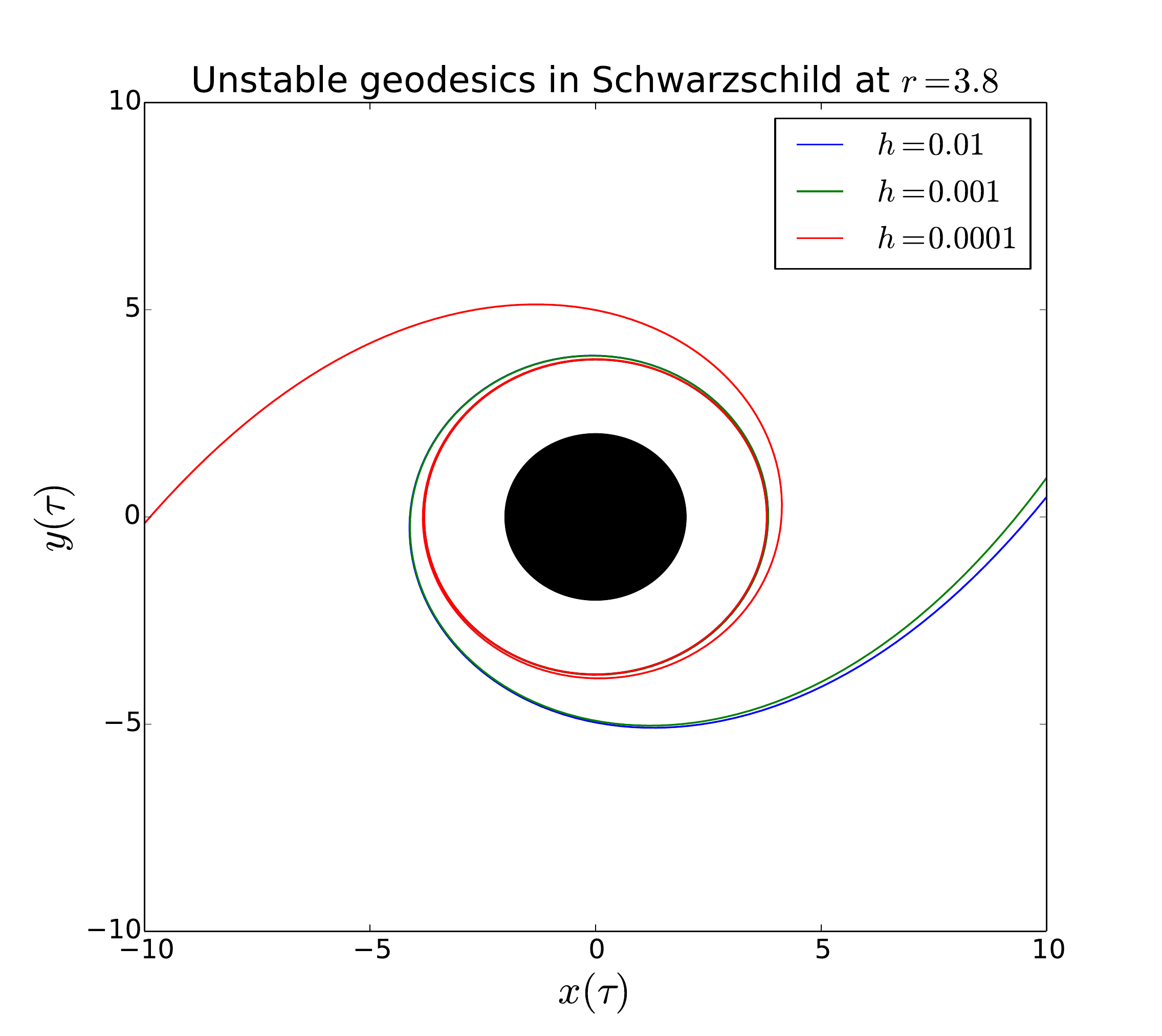}
    \end{minipage}%
    \hfill
    \begin{minipage}[c]{0.5\textwidth}
      \centering
      \includegraphics[width=6.6cm]{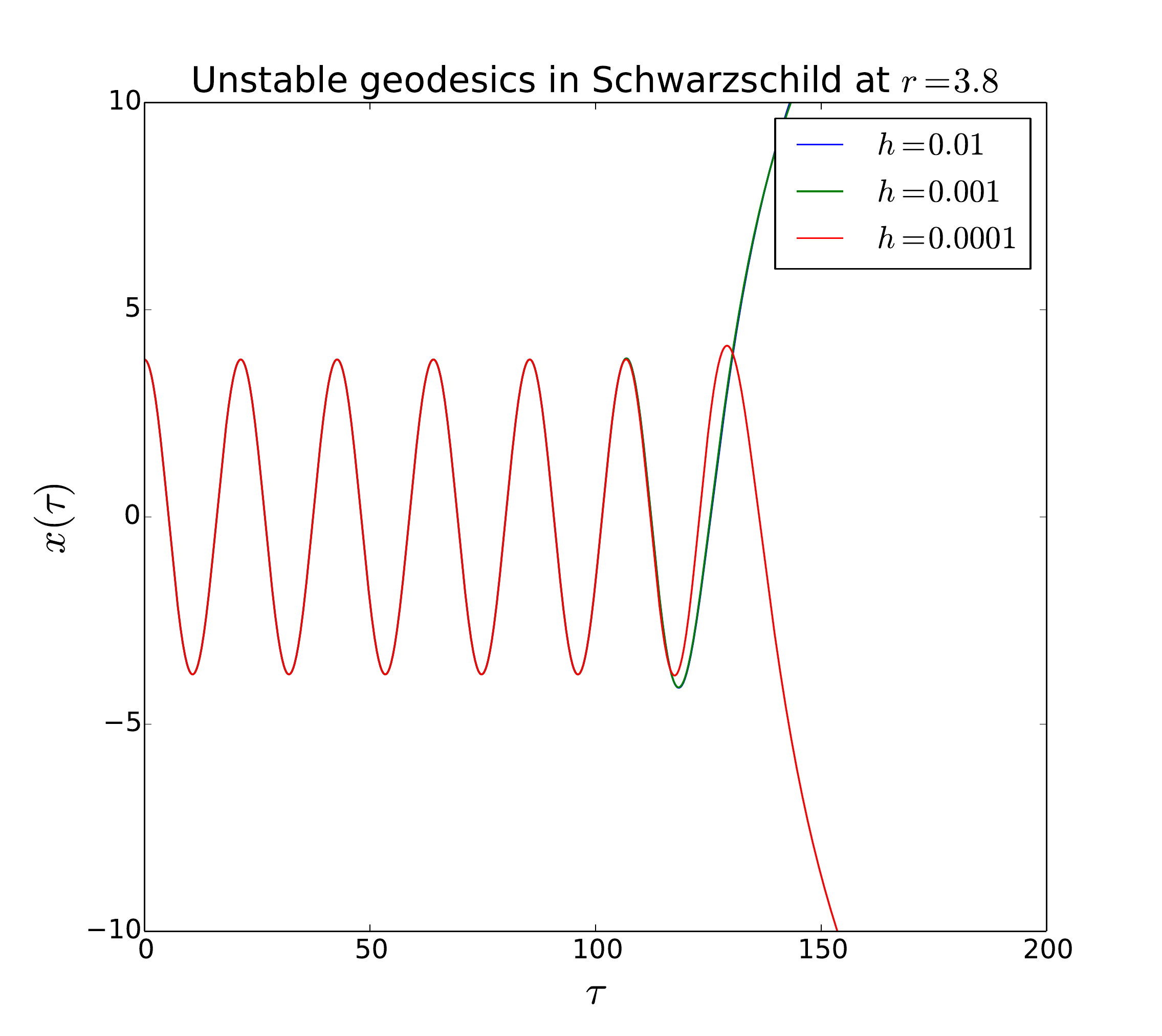}
    \end{minipage}%
  }
   \makebox[\textwidth][\side]{%
    \begin{minipage}[c]{0.5\textwidth}
      \centering
      \includegraphics[width=6.6cm]{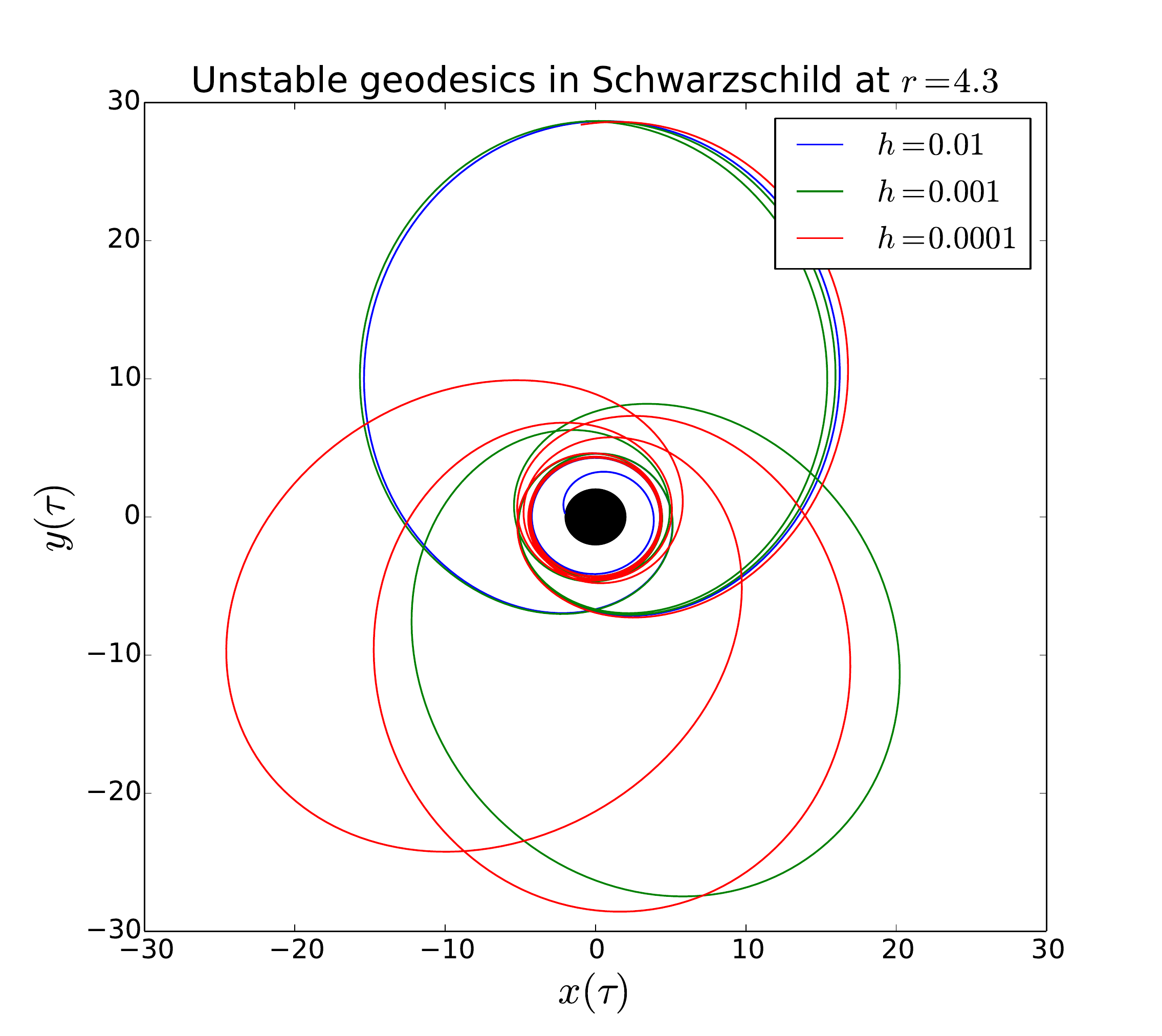}
    \end{minipage}%
    \hfill
    \begin{minipage}[c]{0.5\textwidth}
      \centering
      \includegraphics[width=6.6cm]{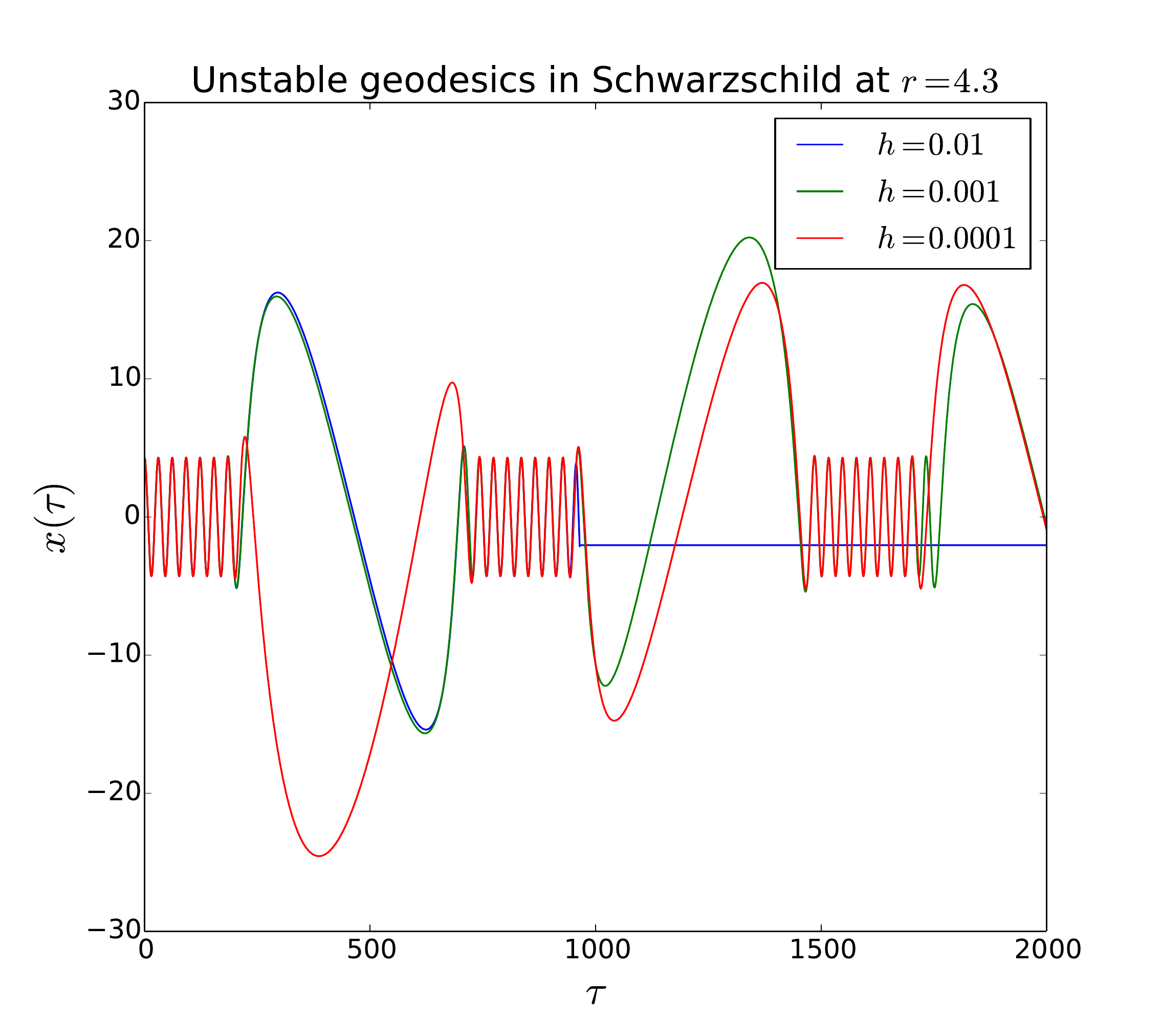}
    \end{minipage}%
  }
\caption{These figures show how the numerical resumption affects the trajectory and the number of oscillations of two unstable geodesics. }
\label{Fig3}
\end{figure}

Consider first, the closed orbits around a Schwarzschild black hole in the equatorial plane. Now, using MALEBC, we compute some unstable orbits in the equatorial plane with a radius between the region $3 \leq r \leq 6$. The fig. \ref{Fig1} show the numerical results of these simulations,
Since these orbits are unstable, they are expected to be sensitive to small fluctuations in the energy and momentum presumably caused by the accumulated error inherent in the numerical method. In other words, these orbits cannot remain in their circular trajectories indefinitely, instead all the unstable trajectories will give a certain number of laps around the compact object before to be expelled or reach the event horizon. On the other hand, the orbits which are located at radius greater and equal to the ISCO, can be kept in their closed circular orbits, as one can see in the fig. \ref{Fig2}.
A different behavior can be observed between the orbits initiated in the unstable region against the stable ones. As mentioned above, small fluctuations in energy and momentum produce that the test particle leaves its circular closed trajectory. This situation can be observed in MALBEC, just running the same initial condition using different integration steps. The figs. \ref{Fig3} show how the step-size $h$, and therefore, the numerical precision (or the numerical resolution) may affect the final state of the test particle.
Note that the blue horizontal line in the figure of the lower right corner represents that the particle reaches the event horizon of the black hole, so the time evolution of the geodesic finished.
On the other hand, stable orbits are more ``robust'' to these differences and remain unchanged regardless of the integration step, as we can see in the following figures,
\begin{figure}[H]
  \checkoddpage
  \edef\side{\ifoddpage l\else r\fi}%
  \makebox[\textwidth][\side]{%
    \begin{minipage}[c]{0.5\textwidth}
      \centering
      \includegraphics[width=6.6cm]{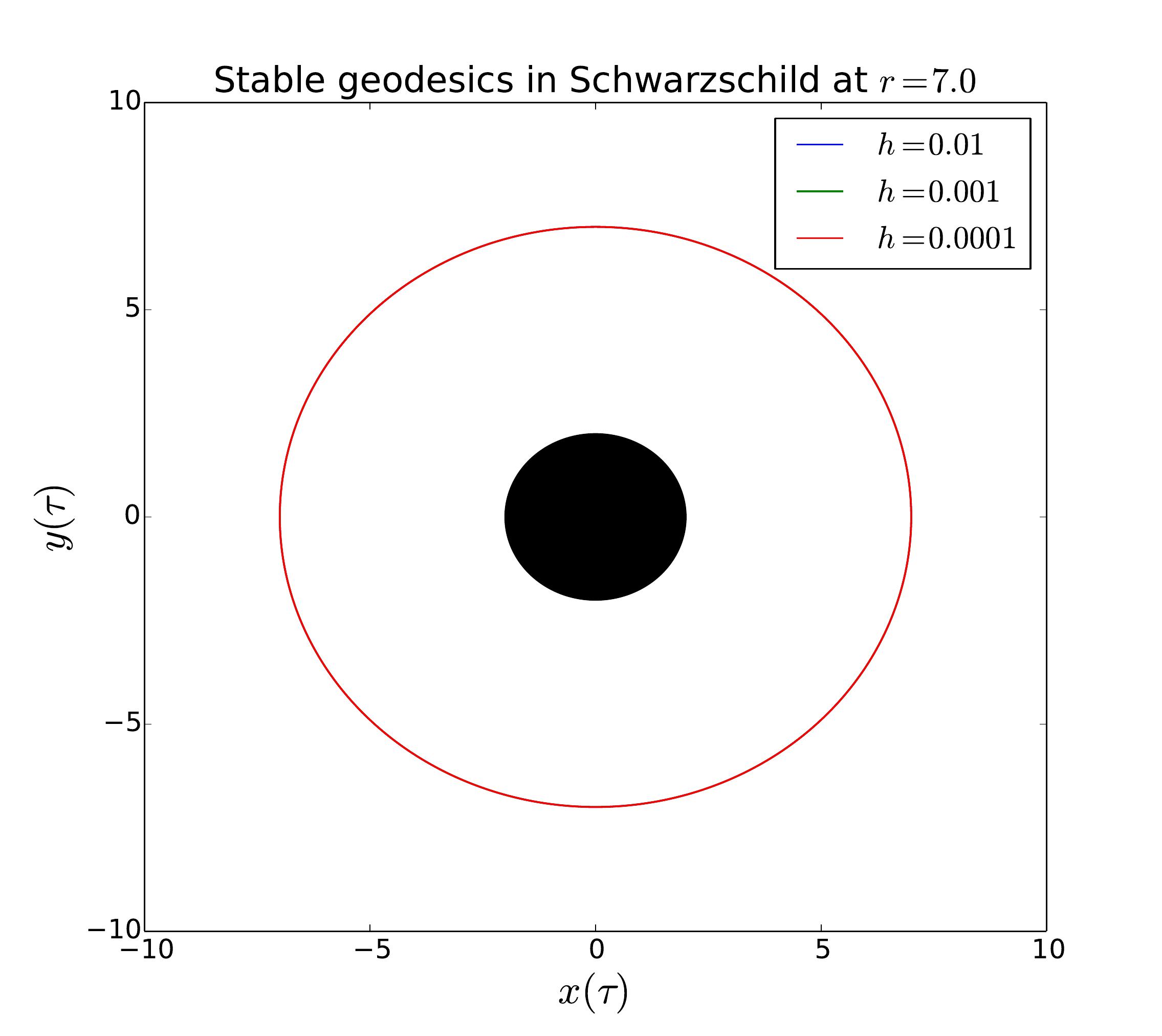}
    \end{minipage}%
    \hfill
    \begin{minipage}[c]{0.5\textwidth}
      \centering
      \includegraphics[width=6.6cm]{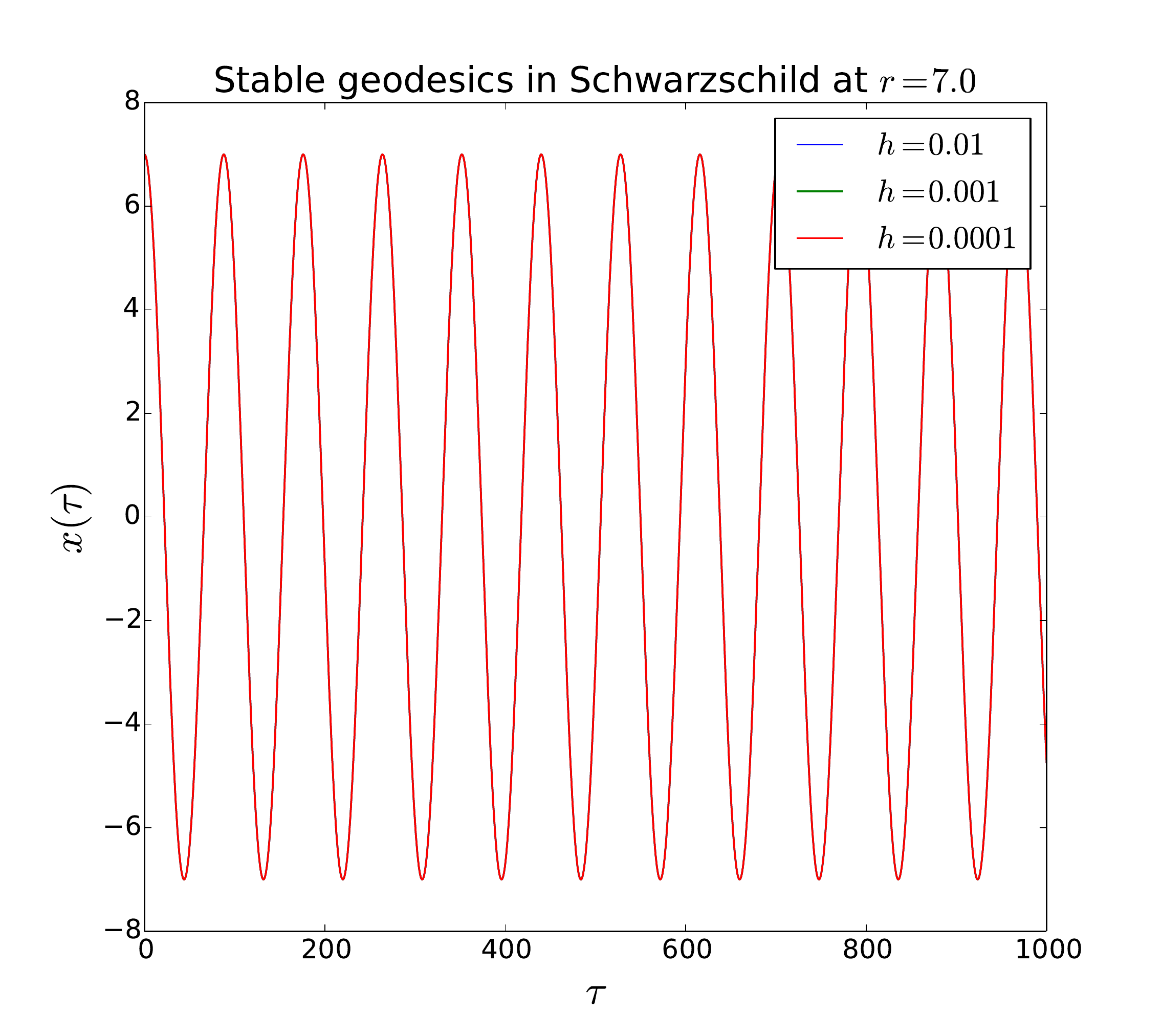}
    \end{minipage}%
  }
\caption{These figures show the trajectory followed by a timelike particle in a stable orbit, the integration was performed using different numerical resolutions.}
\label{Fig4}
\end{figure}
Finally, in these numerical simulations it is possible to evidence the behaviour of the unstable and stable orbits. As one can see in this subsection, the code is able to evolve the timelike test particles by keeping the stable orbits at the same radius around the black hole. However, the geodesics located in the unstable region show the expected variations in the radial coordinate due to the unstable nature of the orbits.

\subsection{Kerr black hole}

\begin{figure}[H]
  \checkoddpage
  \edef\side{\ifoddpage l\else r\fi}%
  \makebox[\textwidth][\side]{%
    \begin{minipage}[c]{0.5\textwidth}
      \centering
      \includegraphics[width=6.6cm]{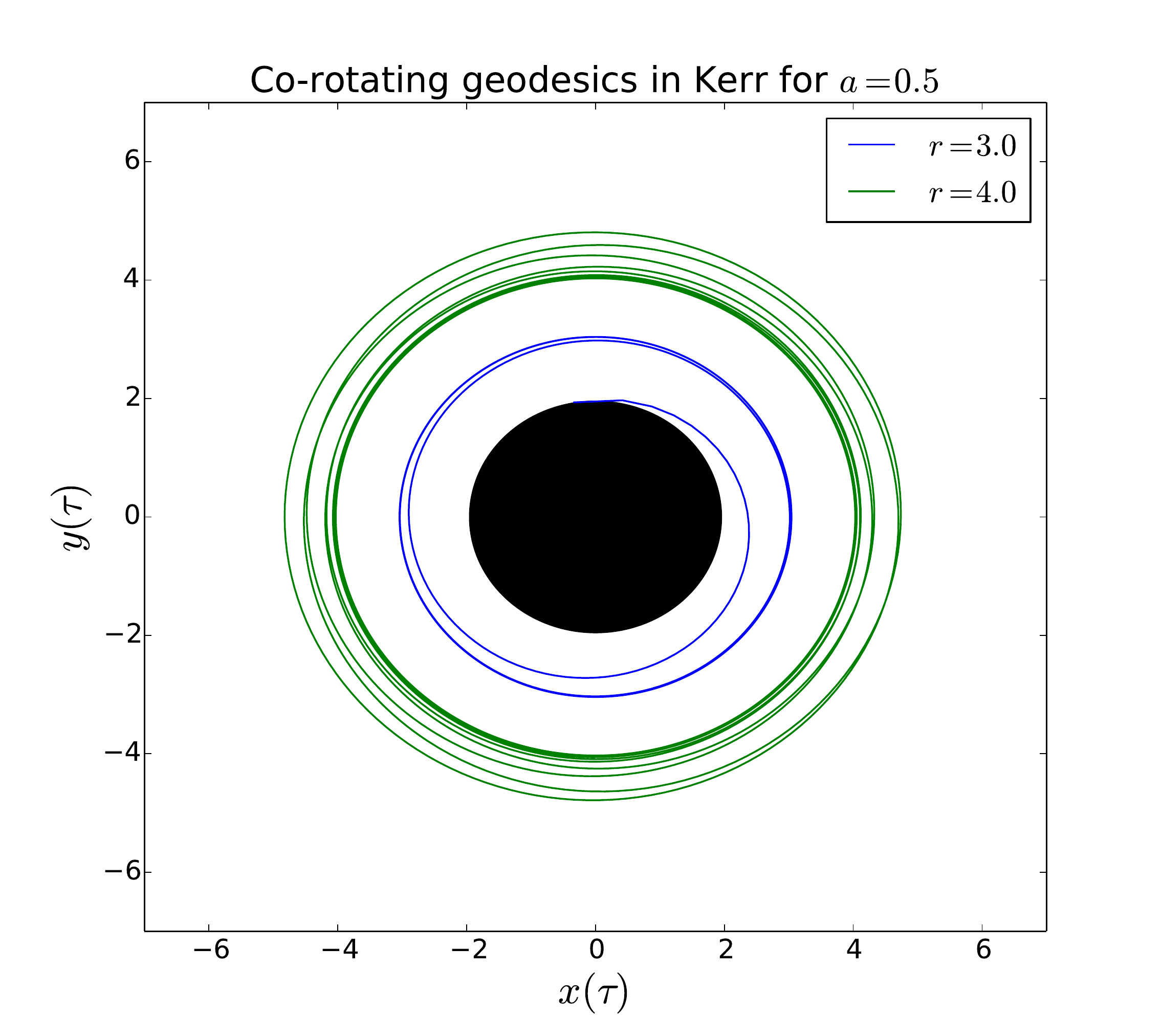}
    \end{minipage}%
    \hfill
    \begin{minipage}[c]{0.5\textwidth}
      \centering
      \includegraphics[width=6.6cm]{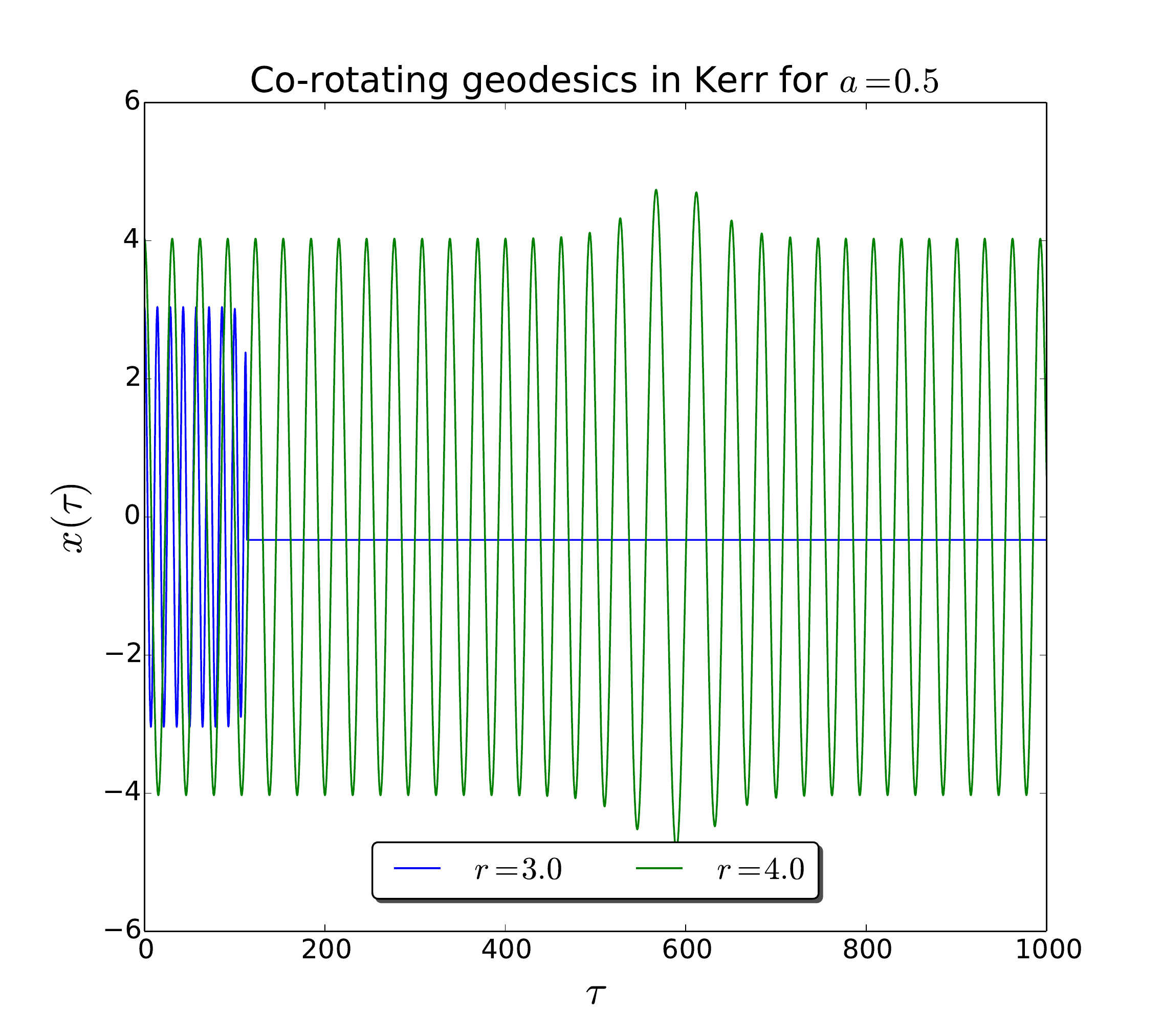}
    \end{minipage}%
  }
   \makebox[\textwidth][\side]{%
    \begin{minipage}[c]{0.5\textwidth}
      \centering
      \includegraphics[width=6.6cm]{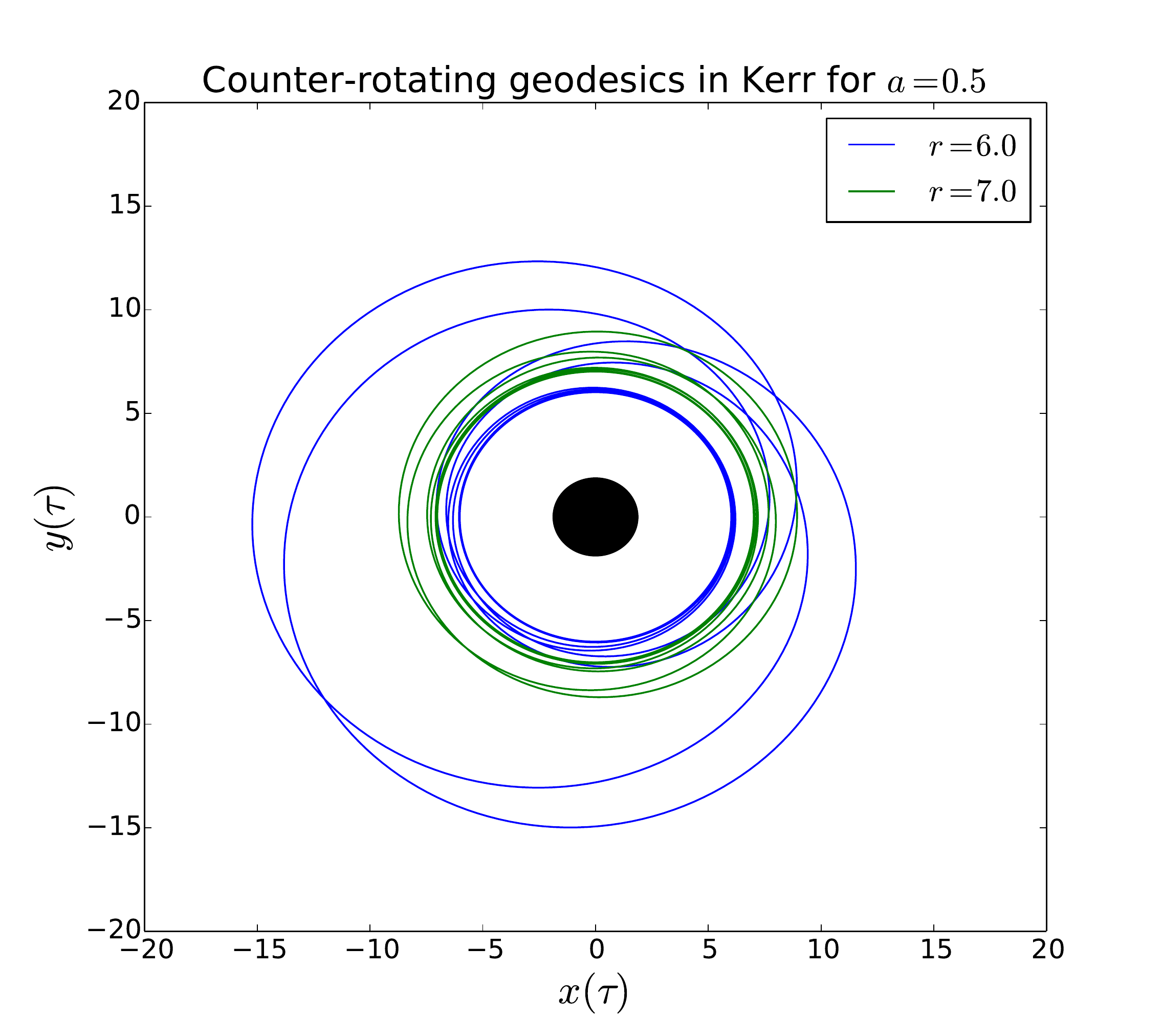}
    \end{minipage}%
    \hfill
    \begin{minipage}[c]{0.5\textwidth}
      \centering
      \includegraphics[width=6.6cm]{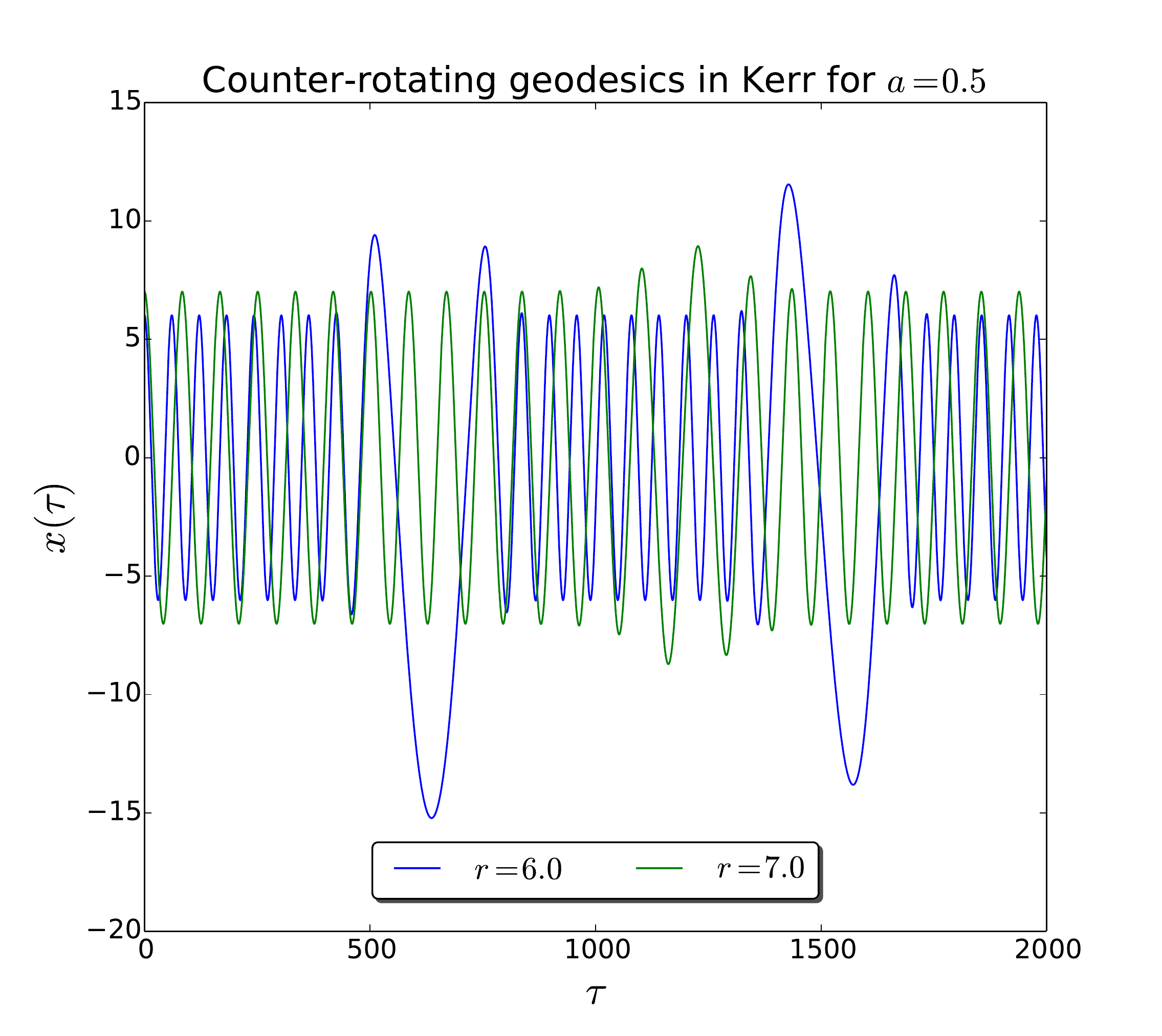}
    \end{minipage}%
  }
\caption{Unstable circular orbits for a Kerr spacetime with $a=0.5$ and $M=1$. The upper figures correspond to co-rotating geodesics inside of the co-rotating ISCO while the inferior are counter-rotating orbits. }
\label{Fig5}
\end{figure}

The Kerr spacetime describes the geometry of a rotating uncharged axially-symmetric black hole, so it is possible to find two types of closed trajectories, one co-rotating to the black hole and the other counter-rotating. All the simulations performed by MALBEC in this subsection will be assuming $M=1$ and Kerr parameter $a=0.5$. However, the results presented here are analogous to those obtained for any $a$. For a Kerr BH of $a=0.5$, the co-rotating and counter-rotating ISCOs are given by the following initial conditions,
\begin{align*}
r_{ISCO}^+&=4.233002529530 \quad p_t^+= 0.9178820066607 \quad p_\phi^+=2.902866153235, \\
r_{ISCO}^-&=7.554584714512 \quad p_t^-=0.9548577730472 \quad p_\phi^-=-3.884212632015,
\end{align*}
where the ISCO radius can be calculated from eq. (\ref{iscokerr}) together with eqs. (\ref{energia}) and (\ref{momento}).
We start by showing some trajectories inside of the ISCO, see fig. \ref{Fig5},
It can be seen from the figures on the right that the test particles oscillate around the compact source with a non-constant radius, also in some cases they can reach the horizon of the black hole. As we can appreciate in the figs. \ref{Fig5}, these particles deviate from their circular trajectories. This kind of behavior is a typical feature of unstable orbits as we discussed in the Schwarzschild subsection. Thus, we can conclude that the co-rotating orbits inside of the co-rotating ISCO, and analogously the counter-rotating orbits inside of the counter-rotating ISCO, are unstable. On the other hand, the geodesics located outside of ISCO are perfectly closed and they keep their radius constant, as we can see in the following figures,
\begin{figure}[H]
  \checkoddpage
  \edef\side{\ifoddpage l\else r\fi}%
  \makebox[\textwidth][\side]{%
    \begin{minipage}[c]{0.5\textwidth}
      \centering
      \includegraphics[width=6.6cm]{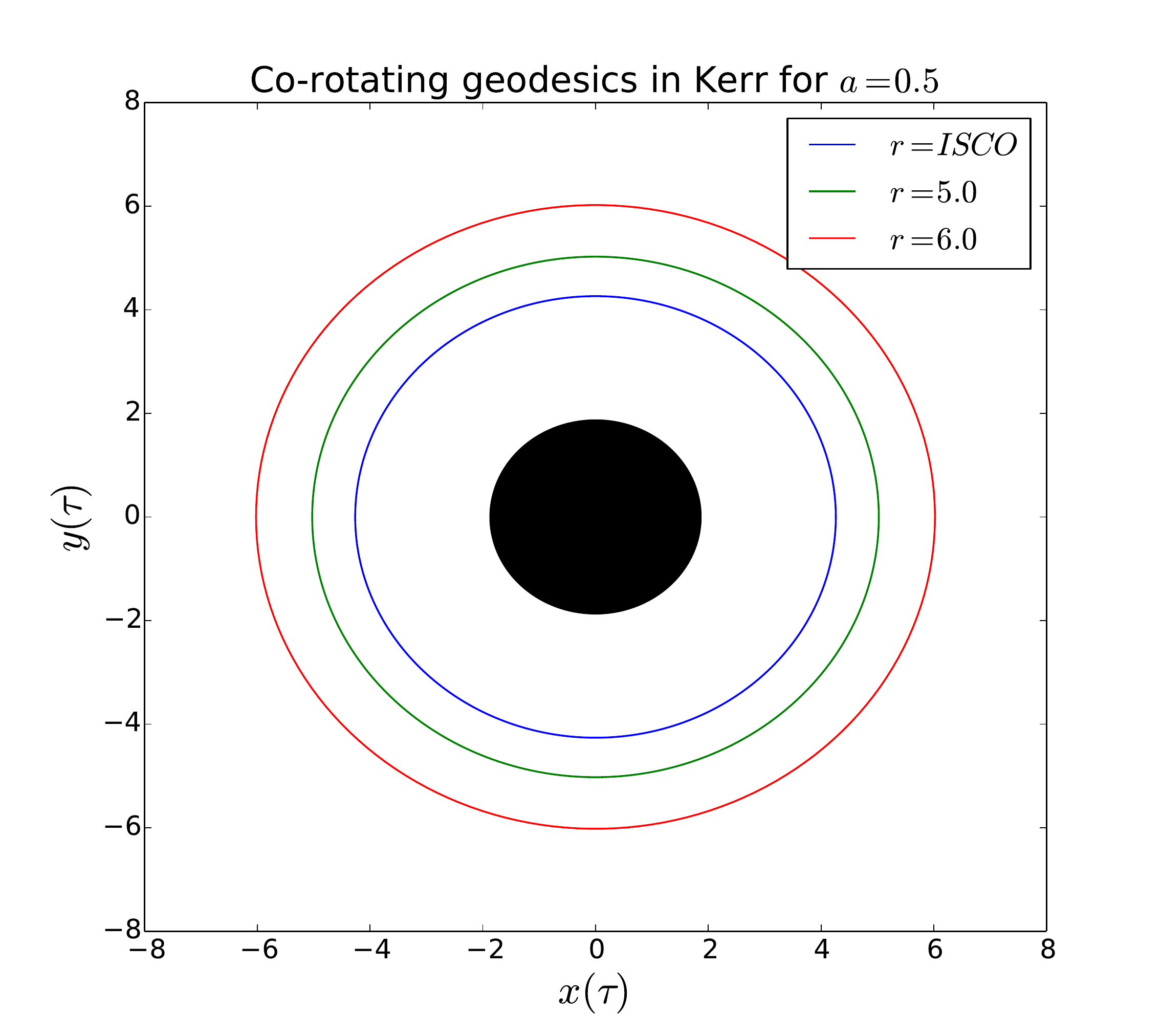}
    \end{minipage}%
    \hfill
    \begin{minipage}[c]{0.5\textwidth}
      \centering
      \includegraphics[width=6.6cm]{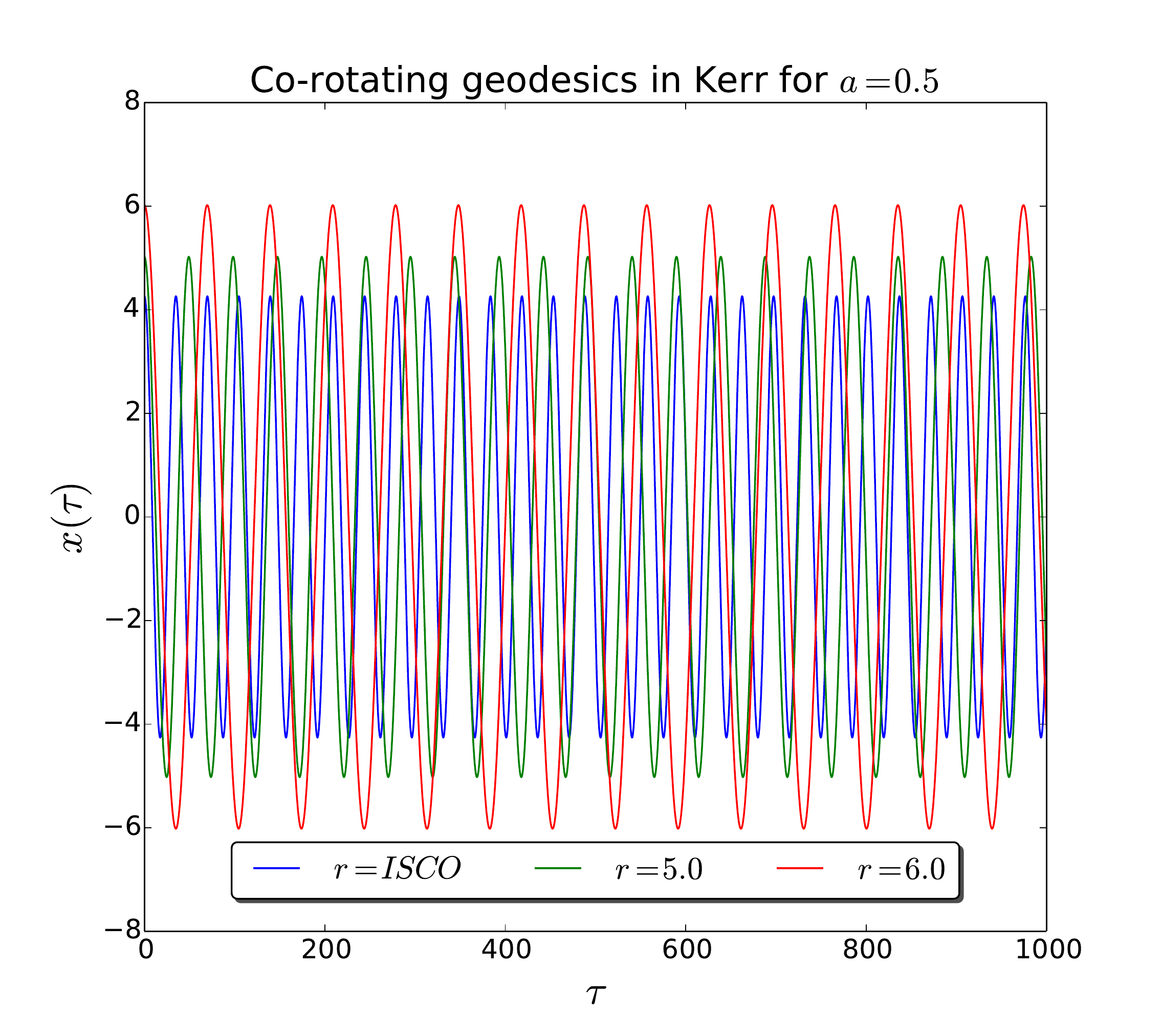}
    \end{minipage}%
  }
   \makebox[\textwidth][\side]{%
    \begin{minipage}[c]{0.5\textwidth}
      \centering
      \includegraphics[width=6.6cm]{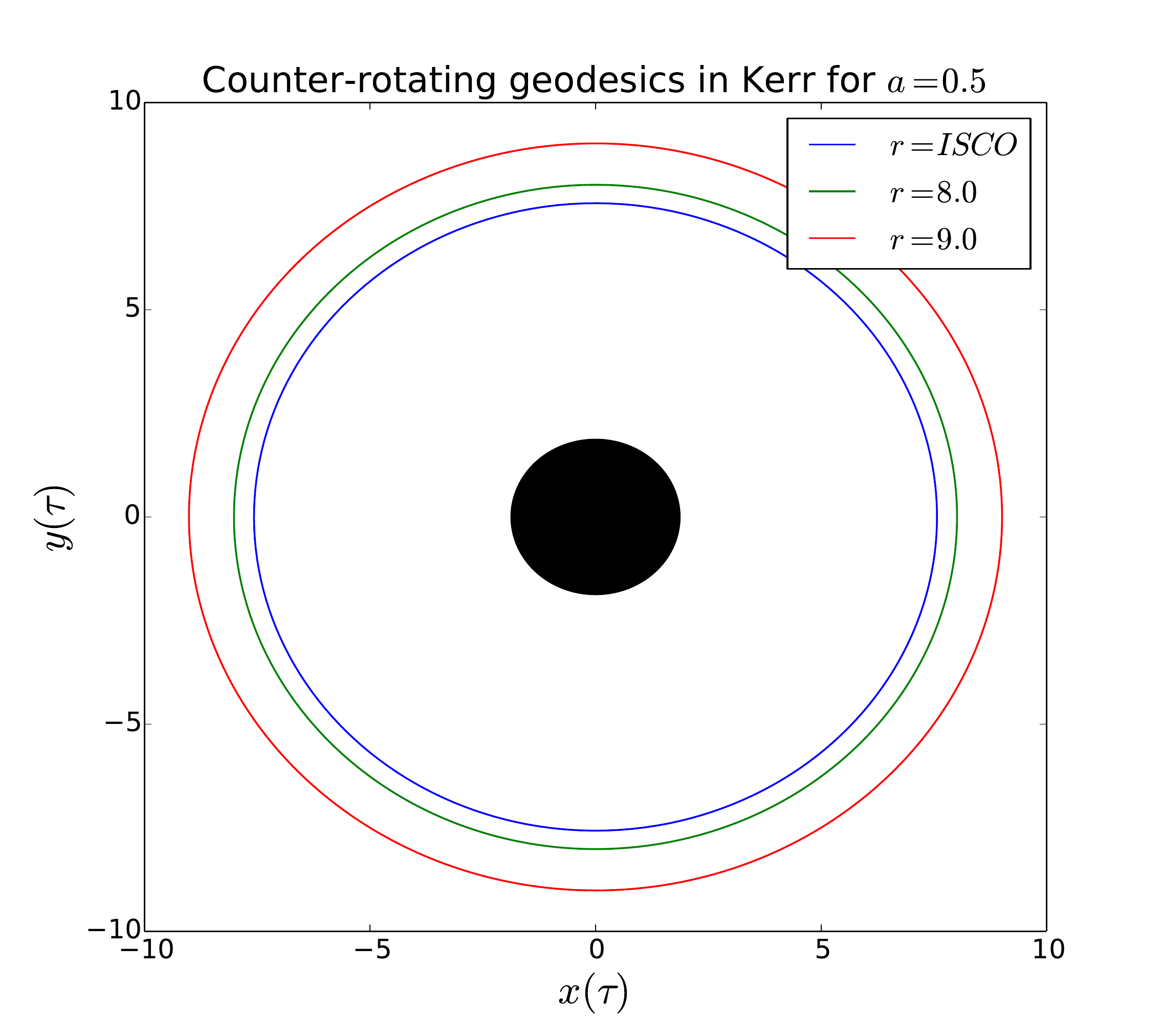}
    \end{minipage}%
    \hfill
    \begin{minipage}[c]{0.5\textwidth}
      \centering
      \includegraphics[width=6.6cm]{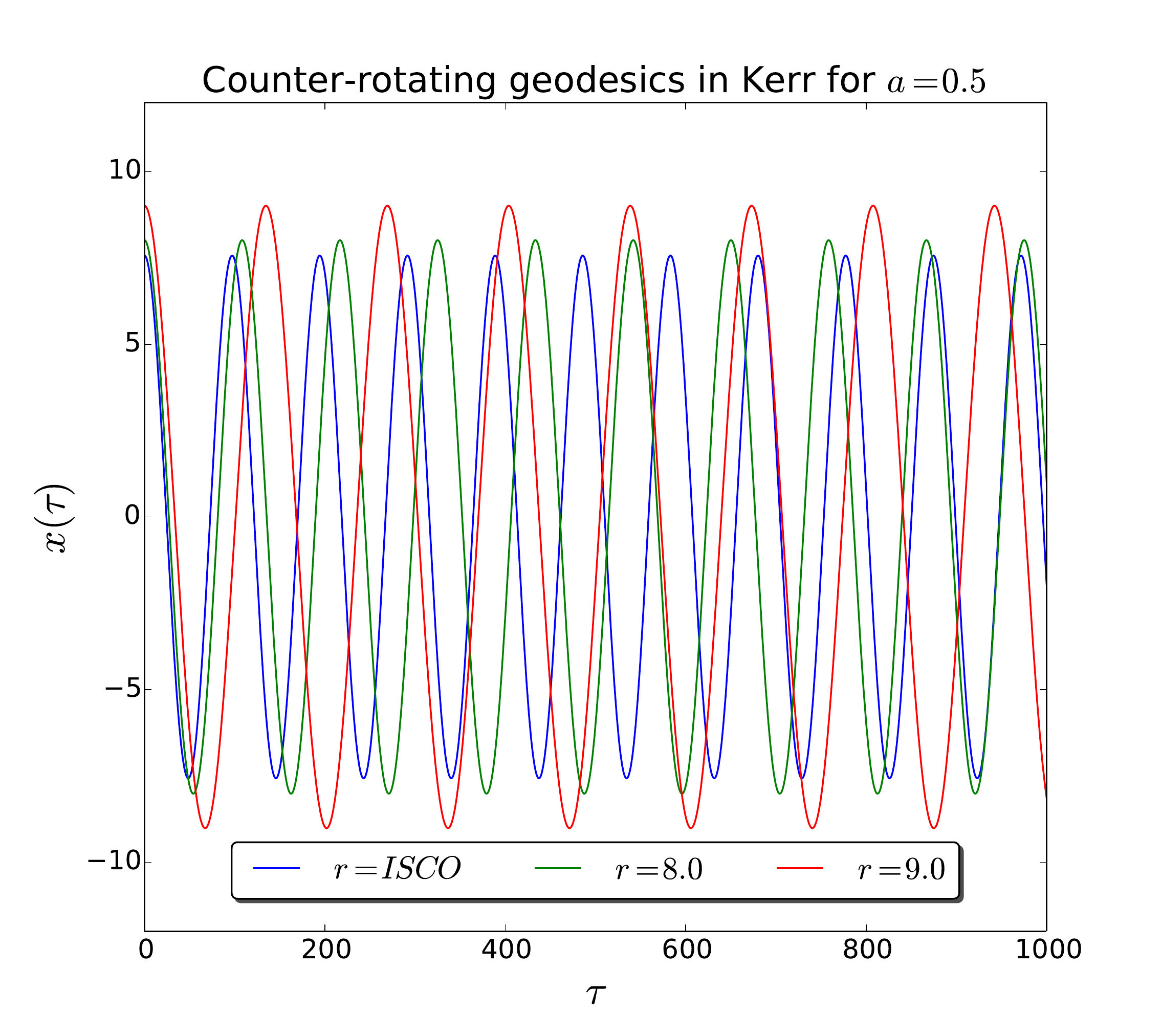}
    \end{minipage}%
  }
\caption{These figures show the ISCO and two stable circular orbits for a Kerr BH of $a=0.5$. The upper figures correspond to the co-rotating trajectories while the lower are the counter-rotating.}
\label{Fig6}
\end{figure}
As in the Schwarzschild case, the orbits initiated in the region $r\geq r_{ISCO}$ are stable trajectories and therefore will not be affected by the level of resolution chosen by the user. Thus, we can highlight two regions, one unstable between the horizon and $r_{ISCO}$, and another stable which is outside of the ISCO. Finally, and for completeness, we compute from eq. (\ref{iscokerr}) the ISCO for several angular momentum-mass ratio for a Kerr BH of unit mass. Firstly, we show the co-rotating ISCOs with their associated Kerr parameter, orbital energy, and angular momentum,
\begin{table}[H]
\centering
\begin{tabular}{|c|c|c|c|}
\hline
$a$ & $r$ & $p_{t}$ & $p_{\phi }$ \\ \hline\hline
0.1 & 5.669302571208 & 0.9393655520586 & 3.367109920810 \\ \hline
0.2 & 5.329443296434 & 0.9353655759711 & 3.264028637021 \\ \hline
0.3 & 4.978616830575 & 0.9306417139417 & 3.153598281507 \\ \hline
0.4 & 4.614335370564 & 0.9249447067862 & 3.034065651040 \\ \hline
0.5 & 4.233002529530 & 0.9178820066607 & 2.902866153235 \\ \hline
0.6 & 3.829069418814 & 0.9087867149047 & 2.755986288640 \\ \hline
0.7 & 3.393128470181 & 0.8963952722476 & 2.586500326227 \\ \hline
0.8 & 2.906643854506 & 0.8778612656718 & 2.380440624350 \\ \hline
0.9 & 2.320883041784 & 0.8442470080056 & 2.099784756124 \\ \hline
1 & 1 & 0.5773502691896 & 1.154700538379 \\ \hline
\end{tabular}
\caption{\label{ISCO-Co} Co-rotating ISCOs in Kerr for $M=1$ and $0.1 \leq a \leq 1$.}
\end{table}
and the counter-rotating orbits are given by,
\begin{table}[H]
\centering
\begin{tabular}{|c|c|c|c|}
\hline
$a$ & $r$ & $p_{t}$ & $p_{\phi }$ \\ \hline\hline
0.1 & 6.322894723789 & 0.9458134386776 & -3.555943626969 \\ \hline
0.2 & 6.639040203567 & 0.9484639574909 & -3.643358857192 \\ \hline
0.3 & 6.949272527004 & 0.9508242198443 & -3.726917098184 \\ \hline
0.4 & 7.254268411283 & 0.9529428538771 & -3.807076678477 \\ \hline
0.5 & 7.554584714512 & 0.9548577730472 & -3.884212632015 \\ \hline
0.6 & 7.850686185306 & 0.9565990449850 & -3.958636355418 \\ \hline
0.7 & 8.142965464834 & 0.9581908703729 & -4.030609693862 \\ \hline
0.8 & 8.431757830806 & 0.9596529818543 & -4.100355267874 \\ \hline
0.9 & 8.717352279606 & 0.9610016543547 & -4.168064196332 \\ \hline
1 & 9 & 0.9622504486493 & -4.233901974057 \\ \hline
\end{tabular}
\caption{\label{ISCO-Count} Counter-rotating ISCOs in Kerr for $M=1$ and $0.1 \leq a \leq 1$.}
\end{table}
These values are useful for testing numerical codes because they delimit the regions of stability of the orbits. Also, these values coincide with those obtained from ref. \cite{jefremov2015innermost} for non-spinning particles. However, it should be noted that our equations are more general in the sense it allows to compute any circular orbit around the compact source, not just the ISCO as in that reference. Finally, based on the numerical results, we can assure that the code has a good error handle, and evolves the system of equations correctly.

\subsection{Numerical Convergence and Error Analysis}
Now, we are ready to investigate about the numerical accuracy of MALBEC by computing the convergence rates using the self-convergence test. For that, we introduce a coefficient $Q$, which is used to measure the convergence order of our RK4 implementation. This coefficient is found by computing the ratio of the differences between  $y_h,y_{h/2},y_{h/4}$, which are the numerical solution with step-sizes $h,h/2$, and $h/4$ respectively. Then, the coefficient $Q$ is defined as,
\begin{equation}
Q=\frac{\parallel y_h - y_{h/2} \parallel }{\parallel y_{h/2} - y_{h/4} \parallel }=2^m, \label{Qt}
\end{equation}
where $m$ is the order of the numerical method, $\parallel y_h - y_{h/2} \parallel$ corresponds to subtraction between the numerical solution $y$ using the step $h$, and $h/2$ respectively. We apply the Euclidean norm to find the differences between $y_h$ and $y_{h/2}$, and also between $y_{h/2}$ and $y_{h/4}$ as follows,
\begin{align}
\parallel y_h - y_{h/2} \parallel&=\sqrt{\sum_{i=1}^8 |y_h^i - y_{h/2}^i|^2},\\
\parallel y_{h/2} - y_{h/4} \parallel&=\sqrt{\sum_{i=1}^8 |y_{h/2}^i - y_{h/4}^i|^2},
\end{align}
here $y_h^i$, $y_{h/2}^i$, and $y_{h/4}^i$ are the i-th components of the numerical solution. The sum ranges from 1 to 8 since we are solving a system of eight ordinary differential equations (see Sec. \ref{Sec2}).

In order to compute the precision coefficient $Q$, we evolve 10 random timelike geodesic in the Kerr spacetime with parameter $a=0.5$. The results of this test is shown in the following figure \ref{figPQ},
\begin{figure}[H]
\centering
\includegraphics[scale=0.4]{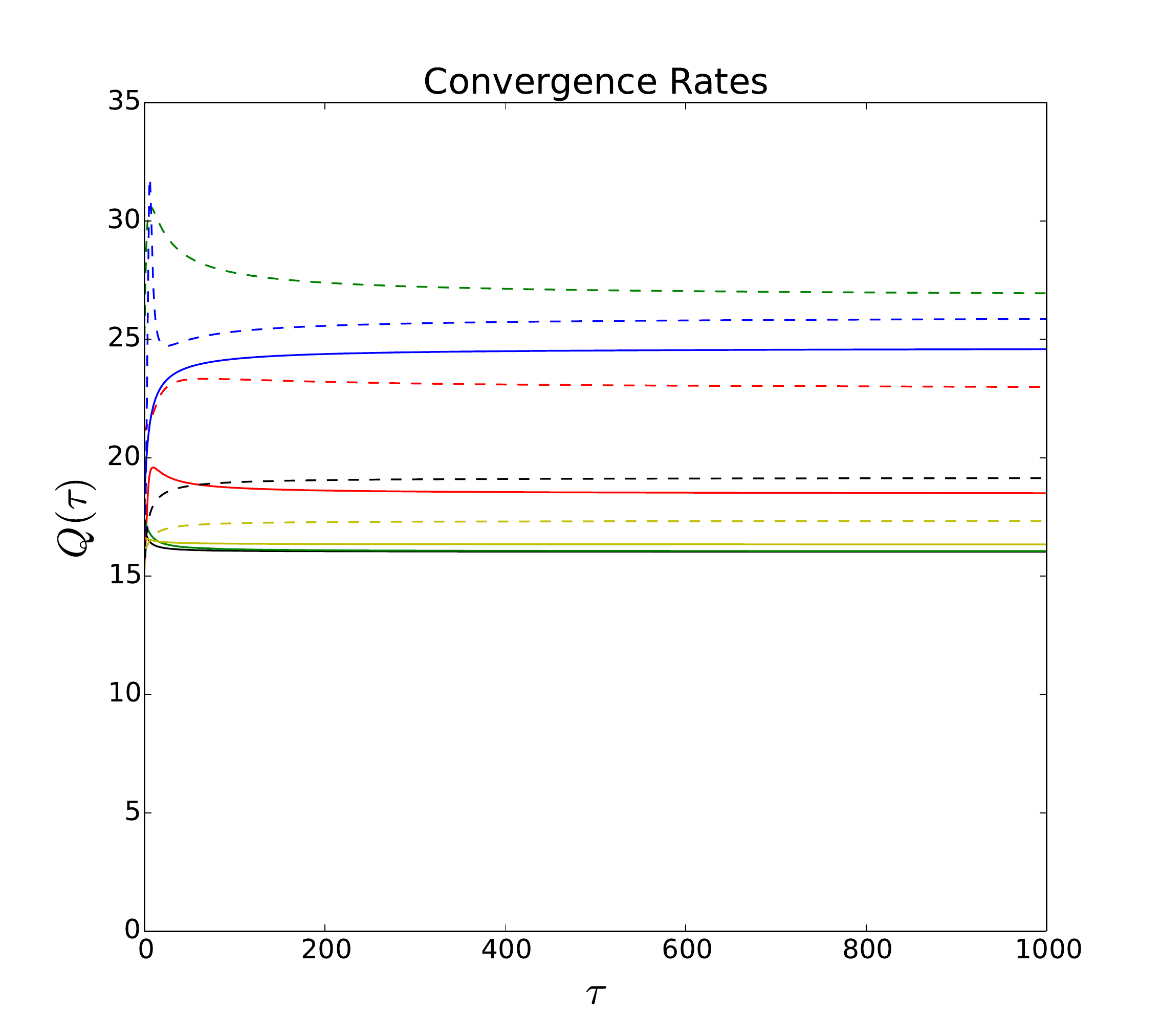}
\caption{\emph{Precision coefficient for 10 timelike geodesics solved numerically in parallel using the RK4 implementation of MALBEC.}}
\label{figPQ}
\end{figure}
Since the RK4 is a fourth order method, the coefficient $Q$ given by eq. (\ref{Qt}) must be greater or closer to $2^4=16$ for all time $\tau$ in a well implemented method. Also is expected that the coefficient $Q$ has fluctuations around 16 or tend to this value as time increases as we can see in the last figure. So, based on the curves plotted in Fig. \ref{figPQ}, where the $Q\geq 16$, we can confirm that our numerical implementation works adequately.

\subsection{Measuring MALBEC performance}

Finally, we focus on evaluating the performance of the Runge-Kutta solver implemented in CUDA. The speed-up of MALBEC will be compared with and equivalent algorithm written in a standard sequential C. All the numerical runs in this paper were performed on the Intel Core i7 CPU with 8 cores running at 2.6 GHz, 8Gb of RAM, a Nvidia Geforce GTX 960M with CUDA Toolkit 7.5.17, and GNU/gcc 4.8.4 installed on the 64-bit GNU/Linux Ubuntu 14.04.
For this performance test, we will measure the runtime needed to solve a set of maximum 50 thousand timelike random geodesics. These geodesics will describe orbits around a Kerr BH with rotation parameter $a=0.5$, the final time for those running will be $t_f=1000$ and the step-size used will be $h=0.1$. We decided to use the Kerr BH instead of Schwarzschild since the system of ODEs is more complex.
The temporal metric used to compute the time during this test is the C function \textbf{clock\_t} located in the \textbf{time.h} header. The time measured in these running are shown in the following table,
\begin{table}[H]
\centering
\begin{tabular}{|c|p{1.5cm}|p{1.5cm}|p{1.3cm}|}
\hline
\backslashbox{Initial Conditions}{Runtime (min)} & $t_{GPU}$ & $t_{CPU}$ & $\frac{t_{CPU}}{t_{GPU}}$  \\\hline
10 & 0.0545956	& 0.04732473 & 0.9 \\\hline
100 & 0.11868735 & 0.49806882 & 4.2 \\\hline
500 & 0.3829293	& 2.43829992 & 6.4 \\\hline
1000 & 0.66657395 & 4.84753887 & 7.3 \\\hline
5000 & 3.543259	& 24.8818684 & 7.0 \\\hline
10000 & 7.14927418 & 48.8400202 & 6.8 \\\hline
50000 & 35.6253648 & 241.872849 & 6.8 \\\hline
\end{tabular}
\caption{\label{table3} Runtime between GPU and CPU for the RK4 implementation.}
\end{table}
Now by plotting these points, it is possible to evidence a linear behavior in the execution times. Therefore, making a linear regression, it is possible to estimate the time necessary to evolve an arbitrary number of initial conditions $n_ic$ in CPU and GPU.
\begin{figure}[H]
\centering
\includegraphics[scale=0.4]{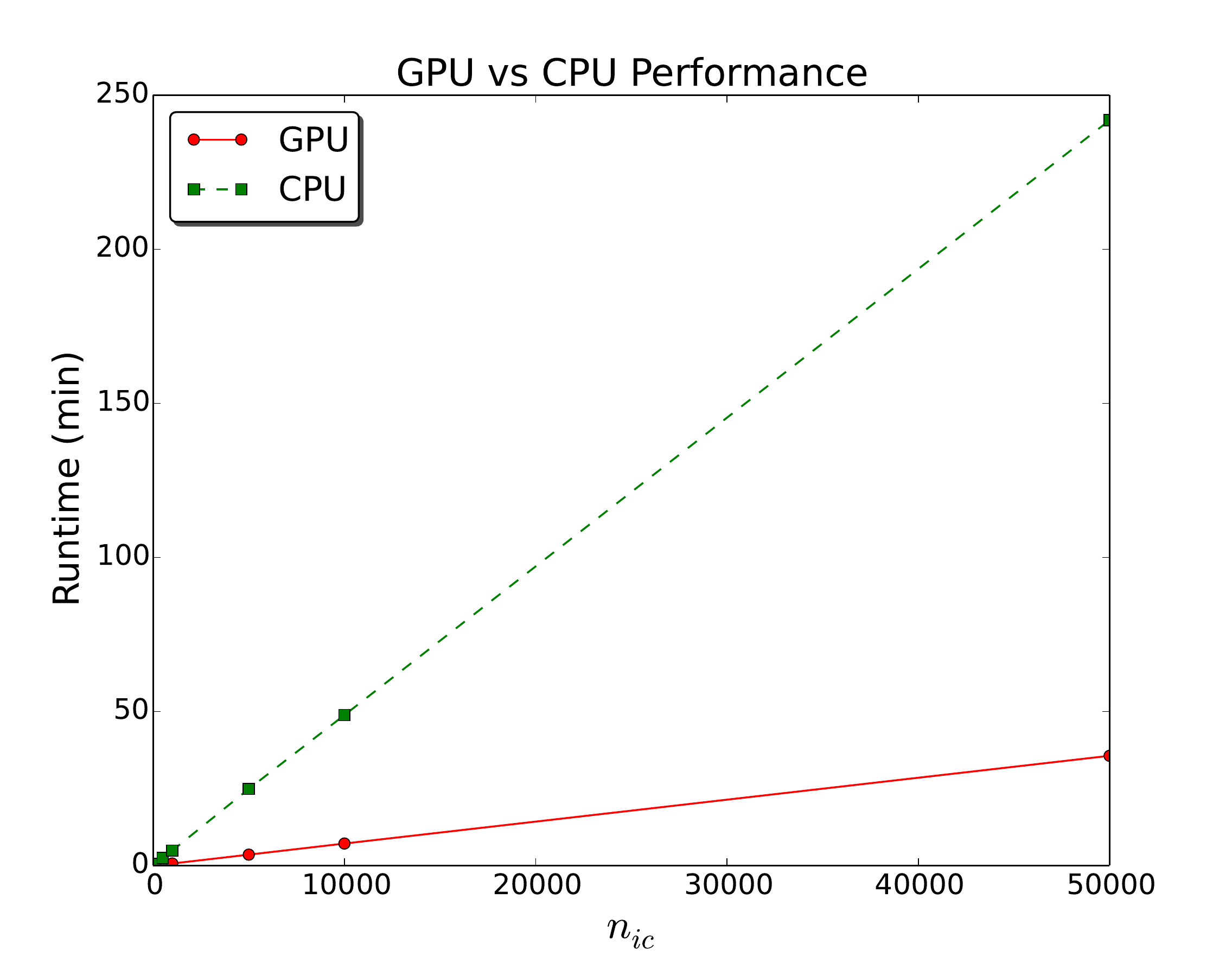}
\figcaption{\emph{The MALBEC runtime when 50 thousand random timelike geodesics in a Kerr spacetime are evolved. The red line correspond to the RK4 GPU implementation while the green line to the CPU. }}
\label{fig1}
\end{figure}
These equations can we written as,
\begin{align}
  t_{GPU} &= 0.0007 n_{ic}+0.0146, \\
  t_{CPU} &= 0.0048 n_{ic} + 0.1876.
\end{align}
Thus, the time needed to run 1 million of initial conditions can be estimated in $t_{CPU}\approx 3.3$ days vs. $t_{GPU}\approx 11.7$ hours, showing an approximate speed-up factor of 7 in our GPU implication.

The speed-up ratio depends mainly on two factors, the GPU-CPU data transfer, and the data write rate to disk. Thus, this factor may increase if the computer used have higher specifications, particularly if the PCI-e bus speed between the device and the host is higher. On the other hand, the data saving can be optimized by using unformatted output, which is the most efficient way to store data, such as a portable binary. In order to illustrate both situations, we will run again a set of initial conditions, but disabling the data  saving in MALBEC and the memory copy from device to host. Finally, the time measured is shown in the following table,
\begin{table}[H]
\centering
\begin{tabular}{|c|p{1.5cm}|p{1.7cm}|p{1.3cm}|}
\hline
\backslashbox{Initial Conditions}{Runtime (sec)} & $t_{GPU}$ & $t_{CPU}$ & $\frac{t_{CPU}}{t_{GPU}}$  \\\hline
10 &	2.16987 &	2.496072 &	1.2 \\\hline
100	& 2.301033 &	26.063659 &	11.3 \\\hline
500	& 4.628749 &	128.347269 &	27.7 \\\hline
1000 &	4.63499 &	254.726977 &	55.0 \\\hline
5000 &	13.598218 &	1270.015683 &	93.4 \\\hline
10000 &	27.066675 &	2550.618381	& 94.2 \\\hline
\end{tabular}
\caption{Runtime between GPU and CPU without data saving.}
\end{table}
As one can see in this table, the performance of the GPU code increases notably while the CPU has a small variation (see table \ref{table3}). MALBEC uses a formatted output, however, the formatted output is more computationally expensive because of the need to convert between internal binary data to ASCII text, besides, the data is written from an array which have to iterate through the array elements. On the other hand, unformatted data is not directly human readable, therefore, special care must be taken when the file is written, since it could happen a mixture between the data of the different orbits evolved simultaneously. We are evaluating to implement the binary outputs, in our next update, in order to improve even more the speed-up of the code.

\section{Final remarks and conclusions}
We introduce and validate a new general relativistic ray-tracer code named MALBEC, this code uses a GPU implementation of the Runge-Kutta solver to integrate null and timelike geodesics around compact sources. The system of ODEs solved by MALBEC is obtained from the 3+1 formulation of the GR, this describe the geodesic motion for test particles moving around Schwarzschild and Kerr black hole.

In order to validate our code, we derive a general set of equations that describe any closed circular timelike orbits around Schwarzschild or Kerr. These equations represent the orbital energy and the angular momentum of the test particle outside the event horizon in the equatorial plane of the source. They are obtained by applying several restrictions, like to assume a constant radius and set the radial momentum to zero in the differential equations that govern the geodesic motion.

These equations are critical to show that MALBEC performs the correct integration of the ODEs system, since the numerical simulation can be contrasted with the theoretical orbits in order to verify the correct numerical evolution. Several orbits, in different stability regions, were subjected to numerical simulations. These simulations show that the test particles fulfil the expected behavior of the theoretical orbits, validating the numerical solutions obtained by MALBEC.

On the other hand, other tests were performed to our CUDA RK4 implementation. First, we check the convergence rates of MALBEC, and then we do a performance test where a large number of simulations were launched in parallel, obtaining a speed-up factor of seven compared with a usual CPU implementation. Finally, based on the excellent results obtained in each test, we can conclude that the first version of MALBEC is ready to be released as a public open source. The code can be downloaded from the following repository: \textcolor[rgb]{0.00,0.07,1.00}{https://github.com/GonzaQuiro/MALBEC}. In a next update, we will include extras metrics and astrophysical objects, we develop a new algorithm to handle numerically computed metrics, and also we investigate several performance technics for increase the speed of MALBEC.

\subsection*{Acknowledgment}
The author  would like to thank to Dr. Omar Ortiz for his suggestions concerning the convergence test.


\end{document}